%% file: imfvar_rev2.tex
\documentclass[useAMS,usenatbib,usegraphicx]{mn2e}

\usepackage{graphicx}
\usepackage[latin1]{inputenc}
\usepackage{color}
\usepackage{times}
\usepackage{natbib}
\usepackage{setspace}
\newif\ifAMStwofonts
\AMStwofontstrue
\definecolor{red}{rgb}{1,0.,0.}

\newcommand{\gaea}{\sc{gaea}}
\def\lesssim{\lower.5ex\hbox{$\; \buildrel < \over \sim \;$}}
\def\gtrsim{\lower.5ex\hbox{$\; \buildrel > \over \sim \;$}}
\title[IGIMF in \gaea] {Variations of the stellar initial mass
  function in semi-analytical models: implications for the mass
  assembly and the chemical enrichment of galaxies in the {\gaea}
  model.}  \author[Fontanot et al.]{
  \parbox[t]{\textwidth}{Fabio Fontanot$^1$\thanks{E-mail:
      fontanot@oats.inaf.it}, Gabriella De Lucia$^1$, Michaela
    Hirschmann$^2$, Gustavo Bruzual$^3$, St\'ephane Charlot$^2$
    and Stefano Zibetti$^4$}
    \vspace*{8pt}\\
    $^1$ INAF - Astronomical Observatory of Trieste, via G.B. Tiepolo 11, I-34143 Trieste, Italy \\
    $^2$ UPMC-CNRS, UMR7095, Institut d'Astrophysique de Paris, 75014, Paris, France \\
    $^3$ Instituto de Radioastronom\'\i a y Astrof\'\i sica, UNAM, Campus Morelia, C.P. 58089, Morelia, M\'exico \\
    $^4$ INAF-Osservatorio Astrofisico di Arcetri, Largo Enrico Fermi 5, I-50125 Firenze, Italy \\
}

\begin{document}
\date{Accepted ... Received ...}

\maketitle

\begin{abstract} 
In this work, we investigate the implications of the Integrated
Galaxy-wide stellar Initial Mass Function (IGIMF) approach in the
framework of the semi-analytic model {\gaea} (GAlaxy Evolution and
Assembly), which features a detailed treatment of chemical enrichment
and stellar feedback. The IGIMF provides an analytic description of
the dependence of the stellar IMF shape on the rate of star formation
in galaxies. We find that our model with a universal IMF predicts a
rather flat [$\alpha$/Fe]-stellar mass relation. The model assuming
the IGIMF, instead, is able to reproduce the observed increase of
$\alpha$-enhancement with stellar mass, in agreement with previous
studies. This is mainly due to the fact that massive galaxies are
characterized by larger star formation rates at high-redshift, leading
to stronger $\alpha$-enhancement with respect to low-mass galaxies. At
the same time, the IGIMF hypothesis does not affect significantly the
trend for shorter star formation timescales for more massive
galaxies. We argue that in the IGIMF scenario the [$\alpha$/Fe] ratios
are good tracers of the highest star formation events. The final
stellar masses and mass-to-light-ratio of our model massive galaxies
are larger than those estimated from the synthetic photometry assuming
a universal IMF, providing a self-consistent interpretation of similar
recent results, based on dynamical analysis of local early type
galaxies.
\end{abstract}


\section{Introduction}\label{sec:intro}
Among the different facets characterising the process of star
formation in galaxies, the shape of the stellar initial mass function
(IMF, defined as the number of stars formed per stellar mass bin in a
given star formation episode) represents an aspect which has not been
fully constrained yet. From a theoretical perspective, a key problem
is the lack of a detailed understanding of the chain of events leading
to the collapse and fragmentation of unstable molecular clouds
\citep{Krumholz14}. On the observational side, direct measurements of
the IMF via stellar counts are possible only in the solar
neighbourhood and/or in the closest galactic systems (i.e. the Milky
Way and its largest satellites). Despite some relevant uncertainties
both at the low-mass-end (in the brown dwarfs regime) and at the
high-mass end (i.e. the exact location of the cutoff), the shape of
the observed IMF shows a remarkable invariance in most Galactic
environments (with the relevant exception of the densest regions of
the Galactic centre, \citealt{Klessen07}). Several functional
representations of the IMF have been proposed in the literature, from
early suggestions of a single power-law \citep{Salpeter55} to a more
recent broken power-law \citep{Kroupa01} and lognormal with a powerlaw
tail \citep{Chabrier03}.

The notion of a {\it universal IMF} has been challenged theoretically
by a number of models exploring the expected impact of small scale
physical properties of the inter-stellar medium (ISM) on the star
formation process \citep[among others]{WeidnerKroupa05, Klessen05,
  HennebelleChabrier08, Hopkins12, Papadopoulos10, Papadopoulos11,
  NarayananDave13}. These models predict a range of possible shapes
for the IMF as a function of the physical properties of star forming
regions, but a direct testing of the range of possible conditions
(i.e. beyond the Local Group environment) is currently impossible, as
we have access only to the integrated light and not to the resolved
stellar populations in distant galaxies. Several indirect
observational evidences for a varying IMF have, however, been reported
in the literature, both in late-type galaxies
\citep{HoverstenGlazebrook08, Gunawardhana11} and in early type
samples \citep{Cappellari12, ConroyvanDokkum12, Ferreras13}, as well
as in dwarf galaxies \citep{McWilliam13}. These claims have raised
considerable debate on the interpretation of the data, and on the
overall consistency between these results \citep[see
  e.g.][]{Smith14}. It is thus timely to test the hypothesis of a
varying IMF in a cosmological context, in order to identify the
constraints coming from the photometric, dynamical and physical
properties of galaxy populations and to correctly interpret the wealth
of data currently available. A variable IMF would indeed impact the
galaxy properties in many aspects, ranging from the chemical
enrichment patterns to the efficiency of stellar feedback, with
critical implications on the fraction of baryonic mass locked in long
lived stars.

Chemical abundance patterns have been used for a long time as an
indication of the star formation timescale of integrated stellar
populations. Indeed, the abundance ratio between $\alpha$ elements (O,
Mg, Si) and iron is critically sensitive to the relative abundance
between short-lived type II core-collapse supernovae (SN, whose main
ejecta are $\alpha$ elements) and long-lived Type Ia SN (the main
iron-peak producers), whose progenitors have lifetimes of the order of
Gyrs \citep[see e.g][and references
  herein]{PipinoMatteucci04}. Assuming a universal IMF, higher levels
of $\alpha$-enhancement require shorter star formation timescales, so
that most of the stars in the system form before the ISM is
iron-enriched by SNIa. The observed increase of [$\alpha$/Fe] ratio
with stellar mass in local early-type galaxies is thus interpreted as
an indication for shorter star formation timescales for massive
galaxies, with respect to their low-mass counterparts
\citep{Matteucci94}. This has been considered a long-standing problem
for theoretical models of galaxy formation and evolution \citep[see
  e.g.][]{Thomas05}. In \citet{Fontanot09b} we analysed the different
definitions of ``downsizing'' trends but could not address explicitly
the abundance ratio trends as a function of stellar mass, which we
dubbed {\it chemoarcheological downsizing}.
\begin{figure}
  \centerline{ \includegraphics[width=9cm]{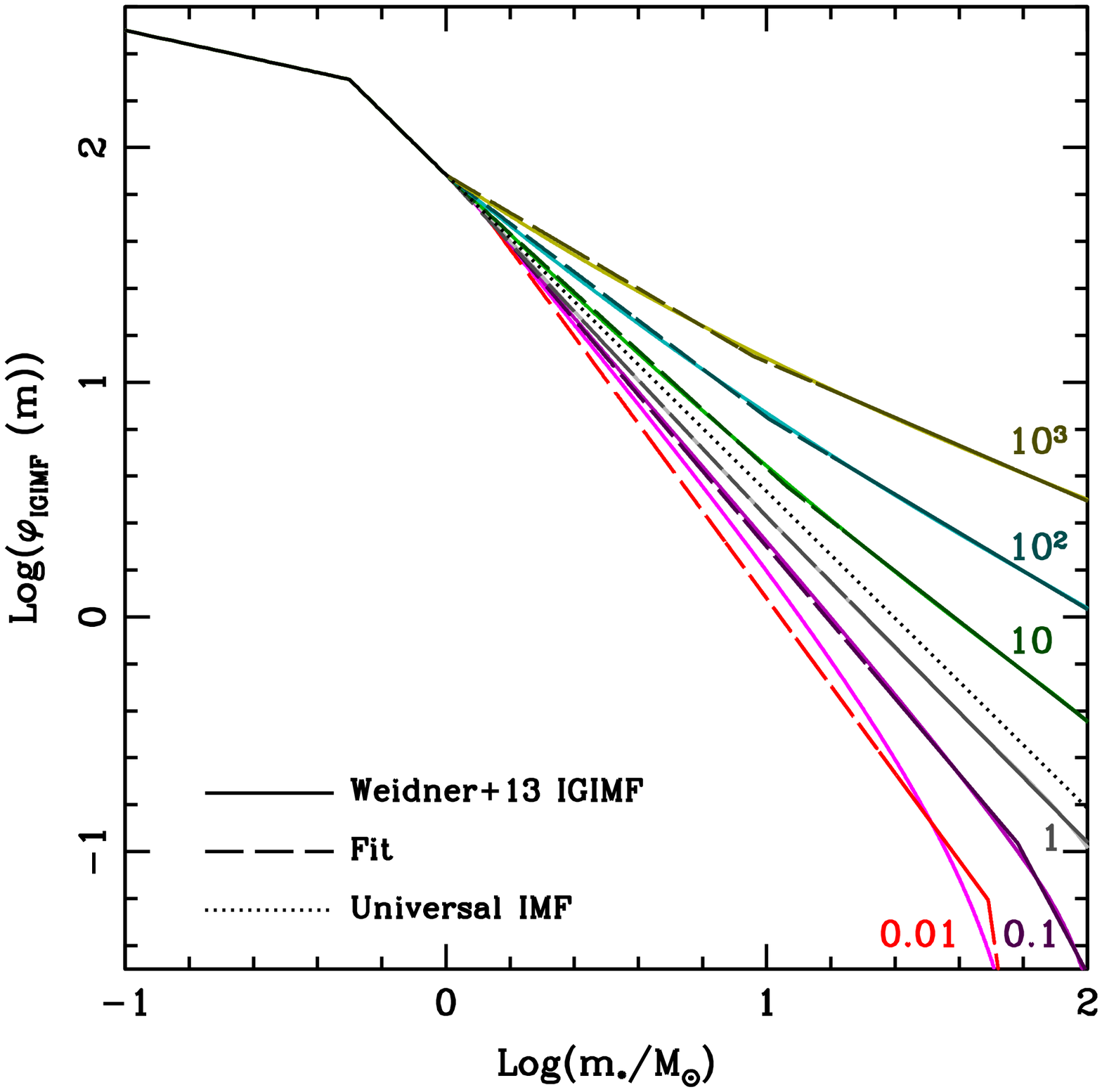} }
  \caption{Integrated galaxy-wide IMF for different star formation
    rates. Each IGIMF is normalised to the same $m_\star<1 M_\odot$
    values. Solid lines correspond to the results of the explicit
    integration using Eq.~\ref{eq:kroimf} to~\ref{eq:rhocl}. Dashed
    lines refer to the four-slopes fits used in our models (see text
    for more details). The universal IMF by \citet{Kroupa01} is shown
    as a dotted line.}\label{fig:igimf}
\end{figure}

Several groups have proposed solutions to this puzzle in the context
of the concordance cosmological model, using both hydrodynamical
simulations and semi-analytical models (SAMs). Two main approaches
have been proposed, mainly focusing on an increased role of feedback
and/or variations of the IMF. \citet{Pipino09} and
\citet{CaluraMenci11} claim that an increased role of AGN feedback (in
the latter model associated with strong stellar feedback in starburst
induced in galaxy interactions) is able to regulate star formation
timescales in the progenitors of massive galaxies. We note that in the
latter model chemical enrichment is computed post-processing star
formation histories extracted from the SAM. Similar results have been
obtained in the framework of hydrodynamical simulations
\citep{Segers16}. Variations of the IMF have also been suggested as
possible drivers of the observed trends. For example,
\citet{Nagashima05} showed that models assuming a Top-Heavy IMF in
starbursts associated with galaxy mergers are in better agreement with
the [$\alpha$/Fe] ratios observed for local elliptical galaxies, than
models using a universal IMF. The first attempt to include a
theoretically based model for a varying IMF in a theoretical model of
galaxy evolution has been presented in \citet[G15
  hereafter]{Gargiulo14}. They implemented the integrated galaxy-wide
IMF (IGIMF) model, first proposed by \citet{WeidnerKroupa05}, in the
Semi-Analytical Galaxies (SAG) model \citep{Cora06}. SAG features a
chemical enrichment scheme which tracks the evolution of individual
chemical elements, taking into account the different timescales
associated with different sources (i.e. SNIa, SNII, stellar winds). In
particular, G15 consider the IGIMF formulation proposed by
\citet{Weidner11}, which relates the shape of the IMF to the star
formation rate (SFR). Their results show that by using a varying IMF
it is possible to recover the positive trend of [$\alpha$/Fe] ratio
with stellar mass, but they did not analyse the implications of this
assumption on the overall assembly and star formation evolution of
model galaxies. Finally, an earlier attempt to compare the predictions
of both theoretically and observationally based IMF variation models
in the SAM framework has been presented in \citet{Fontanot14}.

In this paper, we follow an approach similar to G15. Among the
theoretical models predicting IMF variations tested in
\citet{Fontanot14}, we choose, like G15, to focus on the IGIMF model,
which combines a limited number of physically motivated assumptions in
a set of differential equations, to predict the IMF shape as a
function of the physical properties of the star forming regions. It is
possible to reformulate the key equations as a function of the
(galaxy-wide) SFR, thus providing an elegant formalism, well suited to
be included in a SAM. We thus interface the most recent formulation of
the IGIMF model by \citet{Weidner13} in the GAlaxy Evolution and
Assembly ({\gaea}) model \citep{DeLucia14, Hirschmann16}. As SAG, GAEA
implements a sophisticated model for chemical enrichment, taking into
account the finite lifetimes (and differential yields) of stars of
different mass. On top of that, GAEA also features an updated
formulation for stellar feedback, inspired by numerical simulations,
which allows us to correctly reproduce the redshift evolution of the
galaxy stellar mass function.

This paper is organised as follows. In Section~\ref{sec:igimf} we will
outline the basis for the IGIMF theory as presented in
\citet{Weidner13}. We will then describe its semi-analytic
implementation in Section~\ref{sec:models}. We will present and
discuss our results in Section~\ref{sec:results}. Finally, we will
summarise our conclusions in Section~\ref{sec:final}.

\section{Integrated Galaxy-Wide IMF Theory}\label{sec:igimf}
We compute the integrated galaxy-wide IMF (IGIMF) associated with a
given SFR following the work of \citet[see also \citealt{Kroupa13} for
  a review]{WeidnerKroupa05}. 

Stars form in the densest regions of molecular clouds (MC). The IMF
associated with individual stellar clusters is universal and can be
well represented by a broken power law \citep{Kroupa01}:

\begin{equation}
\varphi_\star(m) =  \left\{
\begin{array}{ll}
(\frac{m}{m_{\rm low}})^{-\alpha_1} & m_{\rm low} \le m < m_0 \\
(\frac{m_0}{m_{\rm low}})^{-\alpha_1} (\frac{m}{m_0})^{-\alpha_2} & m_0 \le m < m_1 \\
(\frac{m_0}{m_{\rm low}})^{-\alpha_1} (\frac{m_1}{m_0})^{-\alpha_2} (\frac{m}{m_1})^{-\alpha_3} & m_1 \le m \le m_{\rm max} \\
\end{array}
\right.
\label{eq:kroimf}
\end{equation}

\noindent
where $m_{\rm low}=0.1$, $m_0=0.5$, $m_1=1.0$, $\alpha_1=1.3$,
$\alpha_2=\alpha_3=2.35$. The shape of the IMF of individual clouds is
usually calibrated on local observations, but it agrees well with
theoretical calculation based on the fragmentation of giant molecular
clouds \citep[see e.g.][and reference herein]{HennebelleChabrier08}.

The key assumption of the IGIMF approach is that the global star
formation activity of the galaxy is well described as the sum over
individual MCs, whose mass function is assumed to be a power-law
\begin{equation}
\varphi_{\rm CL}(M_{\rm cl}) \propto M_{\rm cl}^{-\beta},
\label{eq:clmf}
\end{equation}

\noindent
with local surveys of young embedded clusters suggesting $\beta=2$
\citep{LadaLada03}. The maximum value of the mass of a star cluster
$M_{\rm cl}^{\rm max}$ to form as a function of the instantaneous SFR
has been derived in \citet{Weidner04} using observed maximum star
cluster masses (but it can be derived analytically from optimal
sampling arguments \citealt{Kroupa13}):
\begin{equation}
\log M_{\rm cl}^{\rm max} = 0.746 \log SFR+4.93.
\label{eq:mclmax}
\end{equation}

\noindent
We limit\footnote{Our results do not depend on the exact choice for
  the upper limit cluster mass as Eq.~\ref{eq:mclmax} predicts $M_{\rm
    cl}^{\rm max} > 2 \times 10^7 M_\odot$ only for values of $SFR
  \gtrsim 10^{3.5} M_\odot$ yr$^{-1}$, which never occur in our runs.}
$M_{\rm cl}^{\rm max}$ to $2 \times 10^7 M_\odot$, and the mass of the
smallest star cluster is set to $M_{\rm cl}^{\rm min}=5 M_\odot$
corresponding to individual groups in the Taurus-Auriga complex
\citep{KroupaBouvier03}. At the same time, it is possible to
numerically derive the value of the largest stellar mass ($m_{\rm
  max}$) forming in a cluster, by imposing that it contains exactly
one $m_{\rm max}$ star and using the universal IMF
hypothesis. \citet{PflammAltenburg07} proposed the following fit to
the numerical solution:

\begin{equation}
\begin{array}{ll}
\log m_\star^{\rm max} =& 2.56 \, \log M_{\rm cl} \times \\
    & [ \, 3.82^{9.17} + (\log M_{\rm cl})^{9.17} \, ]^{1/9.17} -0.38.\\
\label{eq:msmax}
\end{array}
\end{equation}

\noindent
Observational data from \citet{Gunawardhana11} require a stronger
flattening of the galaxy-wide mass function slope at large SFRs with
respect to what is inferred from the previous equations. To explain
this result, \citet{Weidner13} assumed that the $\beta$ slope in
Eq.~\ref{eq:clmf} is not universal, but it also depends on SFR:
\begin{equation}
\beta = \left\{
\begin{array}{ll}
2 & SFR < 1 M_\odot/yr \\
-1.06 \log SFR +2 & SFR \ge 1 M_\odot/yr \\
\end{array}
\right.
\label{eq:beta}
\end{equation}
\noindent
Possible variations of the high-mass end $\alpha_3$ of the universal
IMF in individual MCs as a function of cluster core density
($\rho_{\rm cl}$) and/or metallicity have been reported by a number of
authors \citep[see e.g.][and references
  therein]{Kroupa13}. \citet{Marks12} used a principal component
analysis to disentangle among the different possible choices. In the
following, we adopt their proposed dependence of $\alpha_3$ on
$\rho_{\rm cl}$:
\begin{equation}
\alpha_3 = \left\{
\begin{array}{ll}
2.35 & \rho_{\rm cl} < 9.5 \times 10^4 M_\odot/pc^3\\
1.86-0.43 \log(\frac{\rho_{\rm cl}}{10^4}) & \rho_{\rm cl} \ge 9.5 \times 10^4 M_\odot/pc^3 \\
\end{array}
\right.
\label{eq:alpha3}
\end{equation}
\noindent
which has the advantage of being independent of metallicity. It is
possible to theoretically derive the dependence of $\rho_{\rm cl}$ on
$M_{\rm cl}$ \citep{MarksKroupa12}:

\begin{equation}
\log \rho_{\rm cl} = 0.61 \log M_{\rm cl} +2.85
\label{eq:rhocl}
\end{equation}

\noindent
Combining Eq.~\ref{eq:kroimf} to~\ref{eq:rhocl}, the IGIMF
$\varphi_{\rm IGIMF}$ is then defined as \citep[see
  also][]{WeidnerKroupa05}:

\begin{equation}
\varphi_{\rm IGIMF}(m) = \int^{M_{\rm cl}^{\rm max}}_{M_{\rm cl}^{\rm
    min}} \varphi_\star(m \le m_\star^{\rm max} (M_{\rm cl}))
\varphi_{\rm CL}(M_{\rm cl}) dM_{\rm cl}
\label{eq:igimf}
\end{equation}

\noindent 
We stress that in the chosen formulation all the relevant quantities
depend on the value of the SFR, which is the only input quantity
needed from the SAM. The IGIMFs corresponding to 5 different choices
of SFR, normalised to 1 $M_\odot$, are shown in Fig.\ref{fig:igimf}:
while the shape at $m_\star < 1 M_\odot$ is common, large deviations
from the universal $\varphi_\star$ are seen at most SFRs, with the
lowest (highest) levels corresponding to IGIMFs bottom-heavier
(top-heavier) than \citet{Kroupa01}.

Using the theory depicted above it is possible to construct the IGIMF
corresponding to each SFR episode during the evolution of model
galaxies. However, the explicit computation of the IMF at each
integration timestep, and for each model galaxy, would result in a
relevant increase of the computational costs, thus loosing one of the
main advantages of the SAM approach. To avoid this problem, we compute
the shape of IGIMF on a logarithmic grid of 21 SFR values covering the
range $-5< \log SFR <5$ with a 0.5 $dex$ spacing and use the resulting
IGIMF to compute the key quantities needed for our model (see next
section for more details). While running our models, we will thus
assign to each SFR event the IGIMF corresponding to the closest bin in
logarithmic space (if the value lies exactly in between two values, we
will assign the IGIMF corresponding to the lower one).

To better handle the shape of the IGIMF we fit it using a multi
component power-law consistent with an extension of the
\citet{Kroupa01} IMF. In fact, the usual three-slopes approximation
(Eq.~\ref{eq:kroimf}, where $\alpha'_3$ represents the slope fitted at
high masses) is generally not enough for an acceptable fit and we
introduce a fourth slope ($\alpha'_4$) at the high-mass end (which
implies a new break mass $m_{\break}>m_1$). The usual choice $m_1=1$
does not correctly reproduce the position of the second break mass in
our formulation, for $SFR < 1 M_\odot/yr$: in this $SFR$ range we then
treat $m_1$ as a free parameter. Finally, we impose $m_{\rm max}=100
M_\odot$ for consistency with previous studies. The results of the
fitting procedure are shown in Fig.~\ref{fig:igimf} as dashed
lines. The actual parameters for all binned SFR value considered are
listed in Table~\ref{tab:fits}.
\begin{table}
  \caption{Analytical fits to the IGIMF corresponding to different
    star formation rates.}
  \label{tab:fits}
  \renewcommand{\footnoterule}{}
  \centering
  \begin{tabular}{ccccccccc}
    \hline
    $\log SFR$ & $m_1$ & $\alpha'_3$ & $m_{\rm break}$ & $\alpha'_4$ & $m_{\rm max}$ \\
    ${\rm [M}_\odot{\rm /yr]}$ & [M$_\odot$] & & [M$_\odot$] & & [M$_\odot$] \\
    \hline
    -5.0 &   1.200  &  4.422 &   2.276 & 19.474 &   2.64 \\
    -4.5 &   1.288  &  3.817 &   3.914 & 16.923 &   4.72 \\
    -4.0 &   1.288  &  3.408 &   6.574 & 13.926 &   8.36 \\
    -3.5 &   1.286  &  3.179 &  11.087 & 12.227 &  14.79 \\
    -3.0 &   1.288  &  3.038 &  18.754 & 11.063 &  26.05 \\
    -2.5 &   1.287  &  2.937 &  30.972 & 10.412 &  44.16 \\
    -2.0 &   1.287  &  2.861 &  49.022 & 10.421 &  69.61 \\
    -1.5 &   1.287  &  2.766 &  76.816 & 12.991 &  97.87 \\
    -1.0 &   1.288  &  2.614 &  60.522 &  3.551 &   100 \\
    -0.5 &   1.170  &  2.516 &  74.030 &  3.012 &   100 \\
     0.0 &   1      &  2.458 &   9.994 &  2.387 &   100 \\
     0.5 &   1      &  2.354 &  11.774 &  2.214 &   100 \\
     1.0 &   1      &  2.250 &  11.536 &  2.068 &   100 \\
     1.5 &   1      &  2.142 &  10.846 &  1.937 &   100 \\
     2.0 &   1      &  2.033 &  10.137 &  1.817 &   100 \\
     2.5 &   1      &  1.921 &   9.526 &  1.703 &   100 \\
     3.0 &   1      &  1.807 &   9.066 &  1.592 &   100 \\
     3.5 &   1      &  1.731 &   8.833 &  1.538 &   100 \\
     4.0 &   1      &  1.684 &   8.756 &  1.518 &   100 \\
     4.5 &   1      &  1.644 &   8.725 &  1.500 &   100 \\
     5.0 &   1      &  1.609 &   8.746 &  1.485 &   100 \\
    \hline
  \end{tabular}
\end{table}
\begin{figure*}
  \centerline{ \includegraphics[width=18cm]{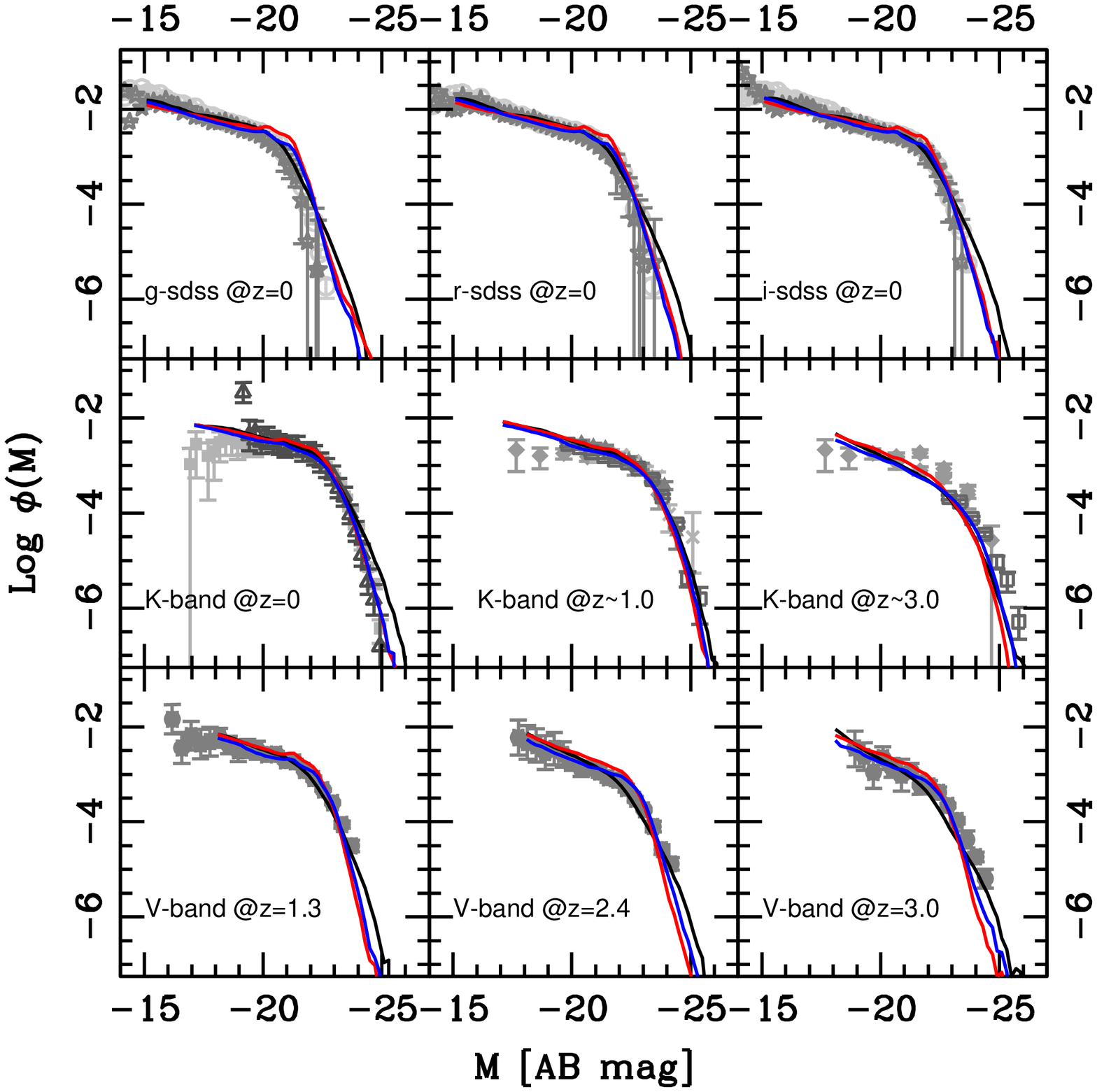} }
  \caption{Predicted luminosity functions in different wavebands and
    at different redshifts. Solid black, red and blue lines refer to
    the predictions from the HDLF16, the Low-$\alpha_{\rm SF}$ and
    High-$\alpha_{\rm SF}$ models respectively. Grey symbols show
    observational estimates in the SDSS $g$, $r$ and $i$-bands by
    \citet[empty circles]{Blanton04} and \citet[stars]{Loveday12}, in
    the $K$-band from \citet[empty triangles]{Kochanek01},
    \citet[filled squares]{Cole01}, \citet[crosses]{Pozzetti03},
    \citet[filled diamonds]{Saracco06}, \citet[empty
      squares]{Cirasuolo10} and in the $V$-band by \citet[filled
      circles]{Marchesini12}. Models have been calibrated to reproduce
    the evolution of the $K$ and $V$-band luminosity
    functions.}\label{fig:lfcal}
\end{figure*}
%
\section{Semi-analytic Model}\label{sec:models}
We test the effect of the IGIMF on the evolution of chemical and
physical properties of galaxies by including it in our semi-analytic
model for GAlaxy Evolution and Assembly ({\gaea}). This model
represents an evolution of that described in \citet{DeLuciaBlaizot07},
and it includes a detailed treatment of chemical enrichment
\citep{DeLucia14} and an improved modeling of stellar feedback
\citep[][HDLF16 hereafter]{Hirschmann16}. Below, we give an overview
of the key ingredients of the model. We refer the interested reader to
the original papers for more details.

Galaxies are assumed to form from gas condensation at the centre of
dark matter haloes, whose evolution is traced using N-body
cosmological simulations. Galaxy evolution results from a complex
network of physical processes including the cooling and heating of
baryonic gas, star formation, accretion of gas onto Super-Massive
Black Holes (SMBHs) and the related feedback processes. In SAMs, these
processes are modelled using analytical and/or numerical prescriptions
that are observationally and/or theoretically motivated. Given the
flexibility and affordable computational costs (with respect to
cosmological hydrodynamical simulations) of this approach, it allows
an efficient sampling of the parameter space, and a quantitative
comparison of its predictions with available observational data.

In the following we will consider predictions from our modified
version of the {\gaea} model. Most of the prescriptions included in
this model are borrowed from \citet{DeLuciaBlaizot07}, with
modifications to follow more accurately processes on the scale of the
Milky Way satellites, as described in \citet{DeLuciaHelmi08} and
\citet{Li10}.

A significant update has been described in \citet{DeLucia14}, who
introduced a new modeling for the chemical enrichment. This new scheme
discards the simplified prescription of instantaneous recycling
approximation and takes into account explicitly the dependence of
stellar evolution on stellar mass. {\gaea} thus traces the evolution of
individual chemical species accounting for finite stellar lifetimes
and differential yields. Briefly, the model assumes stellar lifetimes
parametrizations from \citet{PadovaniMatteucci93}. Stars with
$m_\star<8 M_\odot$ enrich the ISM mainly in their Asymptotic Giant
Branch (AGB) phase: for this population we use the yields from
\citet{Karakas10}. More massive stars are assumed to explode as Type
II SNe releasing metals following the yields by
\citet{ChieffiLimongi02}. The standard {\gaea} model assumes a delay
time distribution for type Ia progenitors corresponding to the single
degenerate scenario of \citet{MatteucciRecchi01} and the
\citet{Thielemann03} metal yields. The probability of a given SNIa
scenario is the only free parameter of the {\gaea} chemical scheme
and, once the chemical yields for individual stars are defined, the
global metal yield of a single stellar population as a function of
time is uniquely predicted by the model (and not treated as a free
parameter).

The second key improvement lies in the updated modeling of stellar
feedback presented in HDLF16. In this paper, different
parametrizations for stellar feedback have been extensively discussed
and compared, with the aim of understanding their impact on the
assembly of galaxies of different stellar mass, in particular with
respect to the delayed formation/evolution of low-mass galaxies
compared to massive ones. Results from HDLF16 show that some form of
preventive or ejective feedback is needed in order to reproduce the
significant evolution of galaxies below the knee of the stellar mass
function. In this paper, we focus on just one of these feedback
schemes, namely the one implementing the scalings derived from the
``Feedback In Realistic Environments'' (FIRE) simulation suite
\citep{Hopkins14}. These ``zoom-in'' hydrodynamic simulations include
sub-grid models that account for individual sources of stellar
feedback (i.e. energy and momentum input from SN explosions, radiative
feedback and stellar winds) and are able to reproduce the baryon
conversion efficiencies at different redshifts. In HDLF16 we adopt the
analytical parametrization for gas reheating proposed by
\citet{Muratov15} as a fit to simulation results. In this
parametrization, the reheating rate depends on both redshift and the
maximum circular velocity of the gas, or only on the stellar mass
(with hardly any redshift dependence). Outflow rates are estimated
adopting the same formulation as in \citet{Guo11}, and gas
reincorporation is modelled as in \citet{Henriques13} assuming an
explicit dependence of the reincorporation time-scale on halo
mass. HDLF16 show that this model is able to reproduce the evolution
of the galaxy stellar mass function, gas fractions and
mass-metallicity relation.

In the following, we will refer to the HDLF16 predictions based on
this feedback scheme as the {\it FIRE} model. For consistency, the
same feedback scheme is adopted also in the IGIMF runs. It is worth
noting here that the FIRE hydrodynamic simulations have been carried
out assuming a universal IMF. This approach does not allow us to
compute in a fully self-consistent way the rate and amount of energy
injected into the inter-stellar medium, which depends on the assumed
IMF. We note, however, that the adopted scalings allow us to reproduce
the observed evolution of the galaxy stellar mass function also in the
framework of the IGIMF theory (see below).

\subsection{Modifications with respect to HDLF16}
\begin{figure}
  \centerline{ \includegraphics[width=9cm]{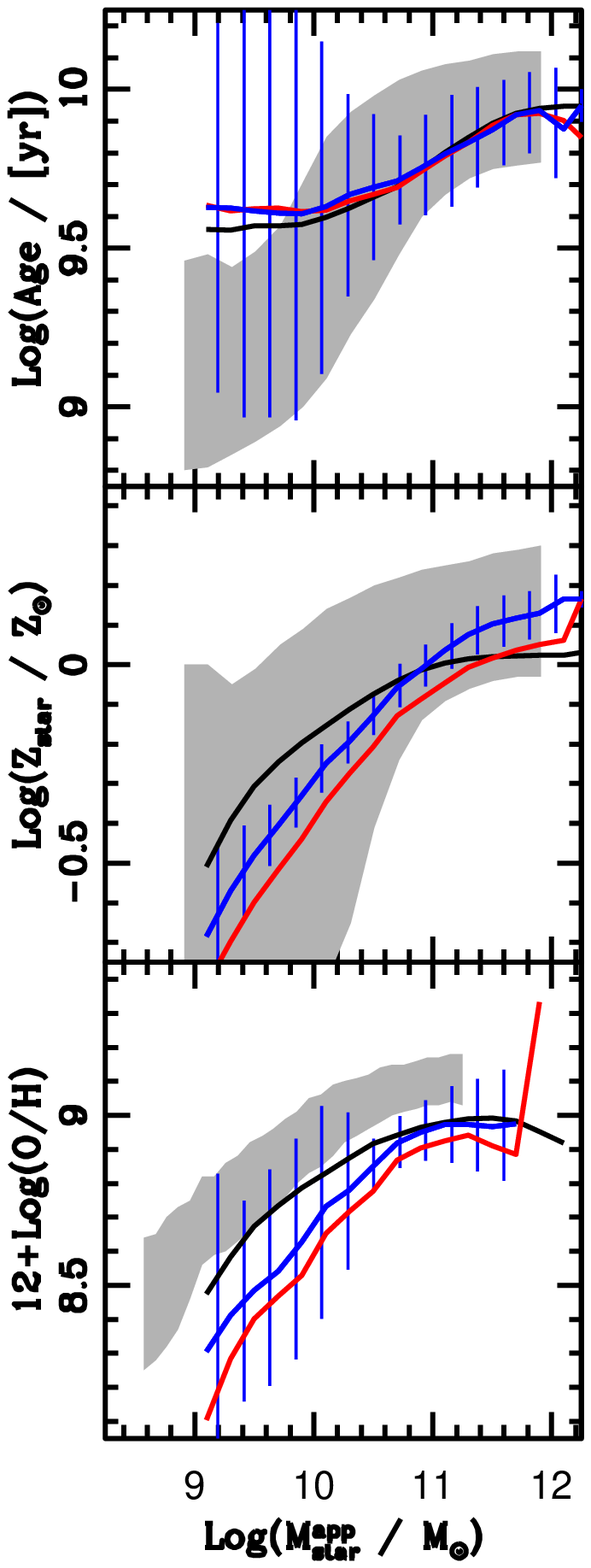} }
  \caption{Local galaxy properties as a function of photometrically
    estimated - Chabrier IMF equivalent - stellar mass $M_\star^{\rm
      app}$: {\it upper panel}: $r$-band luminosity-weighted ages;
    {\it middle panel}: total stellar metallicity; {\it lower panel}:
    cold gas metallicity. Black, red and blue lines represent the mean
    relations from the HDLF16, the Low-$\alpha_{\rm SF}$ and
    High-$\alpha_{\rm SF}$ models respectively, while the hatched
    areas represent the 1-$\sigma$ scatter in the High-$\alpha_{\rm
      SF}$ run. Shaded areas represents observational results from
    \citet{Gallazzi05} and \citet{Tremonti04}, both based on
    SDSS.}\label{fig:mass-met}
\end{figure}

As discussed above, assuming an IGIMF requires associating a
SFR-dependent IMF to each model galaxy, i.e. a different IMF needs to
be selected for each episode of star formation (including those
triggered by mergers). The implementation of this approach in {\gaea}
requires some modifications of the code with respect to the HDLF16
version. The varying shape of the IGIMF affects primarily the baryonic
mass fraction locked in low-mass stars (thus the total stellar mass,
the luminosity and the metallicity of the galaxies), the number of SNe
(thus the strength of stellar feedback), and the different ratio of
Type Ia and Type II SNe (thus the abundance patterns of different gas
phases and stars).

In the framework of the HDLF16 model, the IMF enters in two places:

\begin{itemize}
\item the amount of metals and energy restored into the ISM are
  computed using look-up tables that contain the amount of each
  chemical element considered (including H and He) and energy produced
  by SSP of $1\,{\rm M}_{\sun}$ and distributed according to a
  Chabrier IMF \citep[see][for details]{DeLucia14}. To generalise the
  approach, we construct a suite of tables corresponding to each of
  the IMF bins considered. Each star formation episode (both quiescent
  and merger driven) is then associated to the appropriate table.
\item the photometric properties of galaxies are computed
  interpolating tables containing the luminosity of a single burst of
  fixed mass, as a function of the age and metallicity of the stellar
  population and with a fixed IMF \citep[see][for
    details]{DeLucia04b}. Dust-extinguished magnitudes and
  luminosities are computed using the same approach as in
  \citet{DeLuciaBlaizot07}. As above, we have constructed a set of
  tables corresponding to each IMF bin using an updated version of the
  \citet{Bruzual03} models (G. Bruzual \& S. Charlot, in preparation),
  which include the \citep{Marigo08} prescription for the evolution of
  thermally pulsing AGB (TP-AGB) stars.

\end{itemize}

\subsection{Runs and calibrations}
\begin{table}
  \caption{Parameter values adopted for the runs considered in this study.}
  \label{tab:runs}
  \renewcommand{\footnoterule}{}
  \centering
  \begin{tabular}{cccc}
    \hline
    Parameter & HDLF16 & High-$\alpha_{\rm SF}$ model & Low-$\alpha_{\rm SF}$ model \\
    \hline
    $\alpha_{\rm SF}$            & 0.03 & 0.19  & 0.1 \\
    $\epsilon_{\rm reheat}$       & 0.3  & 0.575 & 0.885 \\
    $\epsilon_{\rm eject}$        & 0.1  & 0.12  & 0.06 \\
    $\gamma_{\rm reinc}$          & 1.0  & 1.0   & 0.68 \\
    $\kappa_{\rm radio} / 10^{-5}$ & 1.0  & 1.78  & 0.87\\
    \hline
  \end{tabular}
\end{table}

In this work, we run {\gaea} on merger trees extracted from the
Millennium Simulation \citep{Springel05}. This simulation assumes a
$\Lambda$CDM concordance model, with parameters derived from WMAP1
(i.e. $\Omega_\Lambda=0.75$, $\Omega_m=0.25$, $\Omega_b=0.045$, $n=1$,
$\sigma_8=0.9$, $H_0=73 \, {\rm km/s/Mpc}$). Although more recent
measurements \citep{Planck_cosmpar} revise these values, we do not
expect the differences in the cosmological parameters to change our
main conclusions, as a minor retuning of the SAM parameters is usually
enough to recover the same level of agreement with data \citep{Wang08,
  Guo13}.

Given the relevant changes induced in galaxy evolution due to the
assumption of a SFR-dependent IGIMF, we recalibrate {\gaea}. The
parameter set used in HDLF16 was tuned on the evolution of the galaxy
stellar mass function. However, most of the estimates for physical
quantities like $M_\star$ and $SFR$ based on photometry and/or
spectroscopy are derived under the assumption of a universal IMF:
therefore they cannot be used for calibrating our version of the SAM
implementing the IGIMF. The only consistent tuning of our model can be
obtained by comparing its predictions with direct (not derived)
observational constraints, i.e. the luminosity functions (LFs).

The observational set used for calibration includes the evolution of
the $K$-band LF \citep{Kochanek01, Cole01, Pozzetti03, Saracco06,
  Cirasuolo10} and the evolution of the $V$-band LF
\citep{Marchesini12} at $z \lesssim 3$. We show the result of the
recalibration procedure in Fig.~\ref{fig:lfcal}, together with the
$z=0$ LFs in the Sloan Digital Sky Survey (SDSS) $g$, $r$ and
$i$-bands \citep{Blanton04, Loveday12}. To recover a good agreement
with the data for the LFs in IGIMF version of {\gaea}, we have to
modify the parameters governing SFR efficiency ($\alpha_{\rm SF}$),
AGN feedback ($\kappa_{\rm radio}$), stellar feedback reheating
($\epsilon_{\rm reheat}$) and ejection rate ($\epsilon_{\rm eject}$),
and the reincorporation rates ($\gamma_{\rm reinc}$).
\begin{figure*}
  \centerline{ \includegraphics[width=18cm]{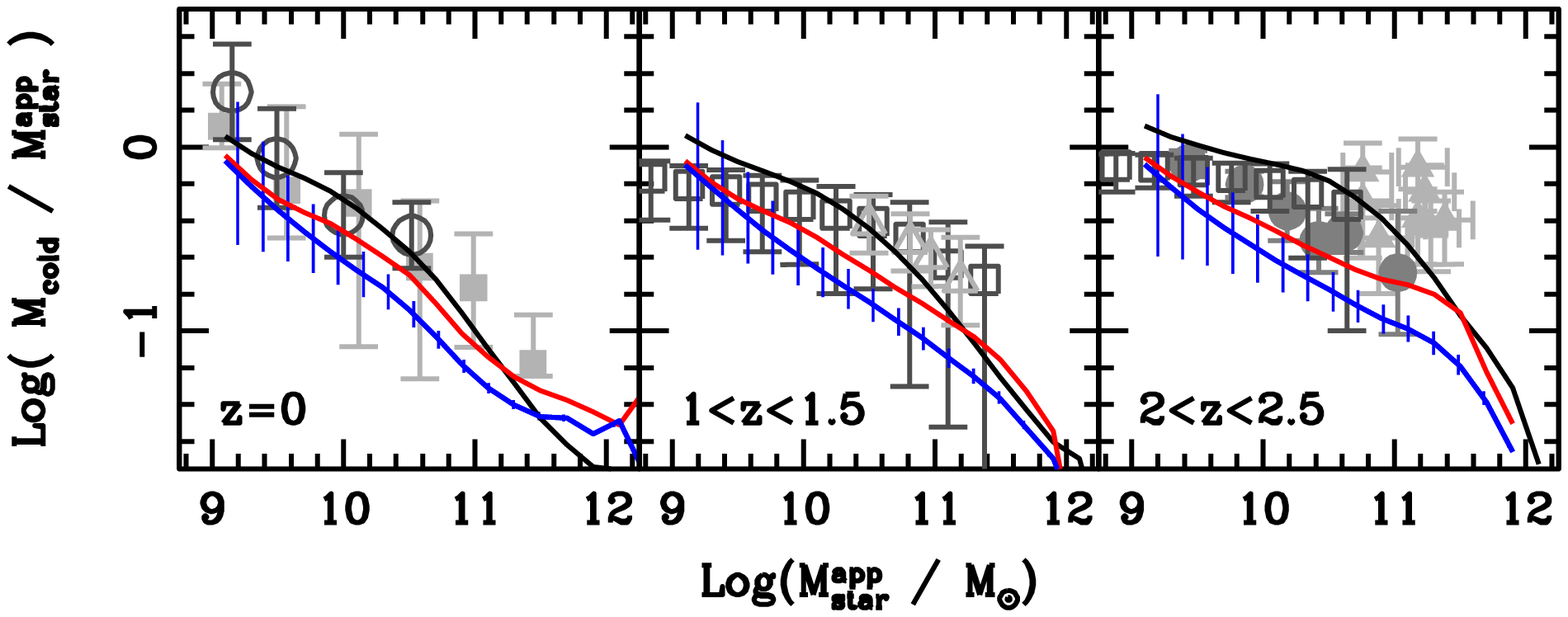} }
  \caption{Redshift evolution of the cold gas fraction ($M_{\rm
      cold}/{M^{\rm app}_{\rm star}}$) of star-forming galaxies
    ($SFR/M_\star^{\rm app}>10^{-2}$ Gyr$^{-1}$). Lines, colours and
    hatched areas represent mean relations as in
    Fig.~\ref{fig:mass-met}. Grey points refer to the data from
    \citet[filled circles]{Erb06}, \citet[filled
      triangles]{Tacconi10}, \citet[empty triangles]{Tacconi13},
    \citet[filled squares]{Peeples14}, \citet[empty
      circles]{Boselli14} and \citet[empty
      squares]{Popping15}.}\label{fig:fgas}
\end{figure*}

We obtain good fits for the chosen LFs in a relatively large area of
the parameter space, typically for star formation efficiencies in the
range from $\sim$10 to $\sim$20 per cent. In the following, we show
predictions for the two extreme runs (High-$\alpha_{\rm SF}$ and
Low-$\alpha_{\rm SF}$), roughly covering this range of star formation
efficiencies. The values of the relevant parameters for these two runs
are listed in Table~\ref{tab:runs}, along with those used in the
HDLF16 model. $\epsilon_{\rm reheat}$ is larger in both IGIMF
realizations than in the reference run. In the Low-$\alpha_{\rm SF}$,
all efficiencies but $\epsilon_{\rm reheat}$ are smaller than the
reference values, while in the High-$\alpha_{\rm SF}$, they are larger
than in HDLF16. No additional calibration on the metal distributions
has been performed with respect to HDLF16, so that all the plots
showing abundance ratios are genuine predictions of the new {\gaea}
version. Whenever stellar masses are estimated from photometry or
spectroscopy (either SED-fitting procedures or colour based scalings),
it could be problematic to compare them with the {\it true} stellar
mass ($M_\star$) predicted by {\gaea}, that depends on the star
formation history of the model galaxy, via the IGIMF theory.

Therefore, we define an {\it apparent} - Chabrier (2003) IMF
equivalent - stellar mass ($M_\star^{\rm app}$) from synthetic
magnitudes using a mass-to-light vs colour relation as commonly done
in the observational literature. We choose to work with the SDSS
$i$-band and $g-i$ as colour, following the results by, e.g.,
\citet[][ZCR09 hereafter]{Zibetti09}. In particular, we adopt the
following relation:

\begin{equation}\label{eq:zibetti}
\log \Upsilon_i = \upsilon (g-i) +\delta
\end{equation} 

where $\Upsilon_i$ represents the stellar mass-to-light ratio in the
$i$-band and $\upsilon=0.90$ and $\delta=0.70$ are best-fit
coefficients derived as in Zibetti et al. (in prep.) from a Monte
Carlo library of 500,000 synthetic stellar population spectra, in a
similar way as in ZCR09. In detail, this library is also based on the
revised version of the Bruzual \& Charlot (2003) SSPs, that includes
an improved treatment of the stellar remnants and of the TP-AGB
evolutionary stage. Dust is treated using the Charlot \& Fall (2000)
2-component prescription, in which enhanced attenuation is applied to
young stars still residing in their birth clouds. With respect to
ZCR09, the new library covers an expanded range of star formations
histories, including raising ones, and implements a simple
prescription for chemical enrichment (a fixed stellar metallicity is
assumed for each model in ZCR09).
As a consistency check we compare $M_\star^{\rm app}$ and $M_\star$
for the HDLF16 run with universal IMF. The two quantities are tightly
correlated, but $M_\star^{\rm app}$ shows a constant shift with
respect to $M_\star$ of the order of 0.1 dex. ZCR09 show that a
similar shift may be explained by spatial resolution effects:
$M_\star$ estimated from integrated photometry is systematically lower
than the stellar mass obtained from resolved photometry because
younger and less dust-obscured regions dominate the light and bias
colours blue, hence the mass-to-light ratios low. Although ZCR09 does
not have a statistical sample for a quantitative assessment of this
effect, preliminary results from a sample of a few hundreds
CALIFA\footnote{Calar Alto Legacy Integral Field spectroscopy Area
  survey} \citep{Sanchez12, Walcher14} galaxies confirm the effect
with an amplitude very similar to the one found here. Therefore, in
all runs we compensate for this by adding to $\delta$ an additional
shift of $0.13$. This formulation implies that using $M_\star^{\rm
  app}$ or $M_\star$ is equivalent in HDLF16 (in a statistical sense,
i.e. modulo some scatter).

\section{Results \& Discussion}\label{sec:results}

\subsection{Basic Predictions}

In Fig~\ref{fig:mass-met}, we show the relations between the
luminosity-weighted stellar age (top panel), stellar metallicity
(middle panel), cold gas metallicity (bottom panel) and stellar
mass. As the stellar masses used in the observational relations
\citep{Tremonti04, Gallazzi05} are based on photometric estimates
(e.g. SED-fitting), we use our apparent stellar masses $M_\star^{\rm
  app}$ for the comparison. In all panels in Fig.~\ref{fig:mass-met},
the black solid line refers to the predictions of the HDLF16 model,
while the red and blue lines to the predictions of our IGIMF-based
{\gaea} realizations. Qualitatively, the IGIMF does not affect
significantly any of the considered trends. The $z=0$ mass-metallicity
relations tend to be somewhat steeper for the IGIMF runs than in the
reference model, which is mainly due to low-mass galaxies being less
enriched. It is worth stressing that the normalization of the observed
mass-(cold gas) metallicity relation depends on the metallicity tracer
used and the overall consistency among different estimates is of the
order of $\sim$ 0.2 dex \citep{KewleyEllison08}. Considered that
\citet{Tremonti04} use a tracer biased in favour of high oxygen
abundances, our predictions are still compatible with the data. The
very steep relation found in the IGIMF runs is a potential issue, as
the \citet{Tremonti04} mass-metallicity relation is the steepest among
the various observational determinations. We defer a more detailed
analysis of the mass-metallicity relation (and its
redshift-evolution), to a future work (De Lucia et al., in
preparation).

Another relevant test for our IGIMF runs involves the amount of cold
gas associated with model galaxies. In Fig.~\ref{fig:fgas} we show the
redshift evolution of the cold gas fraction ($M_{\rm cold}/{M^{\rm
    app}_{\rm star}}$) of star-forming galaxies (defined as
$SFR/M_\star^{\rm app}>10^{-2}$ Gyr$^{-1}$). At fixed $M_\star^{\rm
  app}$, both IGIMF runs predict gas fractions that are systematically
lower than those predicted by the HDLF16 model, with a clear trend for
decreasing fractions at increasing $\alpha_{\rm SF}$. This behaviour
is expected, as the larger SFR efficiencies in the IGIMF runs
correspond to stronger reheating efficiencies, which deplete the cold
gas more effectively. The Low-$\alpha_{\rm SF}$ run formally provides
the best-fit to the data; however, none of the models is ruled out by
the available data, as the observational uncertainties are large.

\subsection{[O/Fe] ratios.}
\begin{figure}
  \centerline{ \includegraphics[width=9cm]{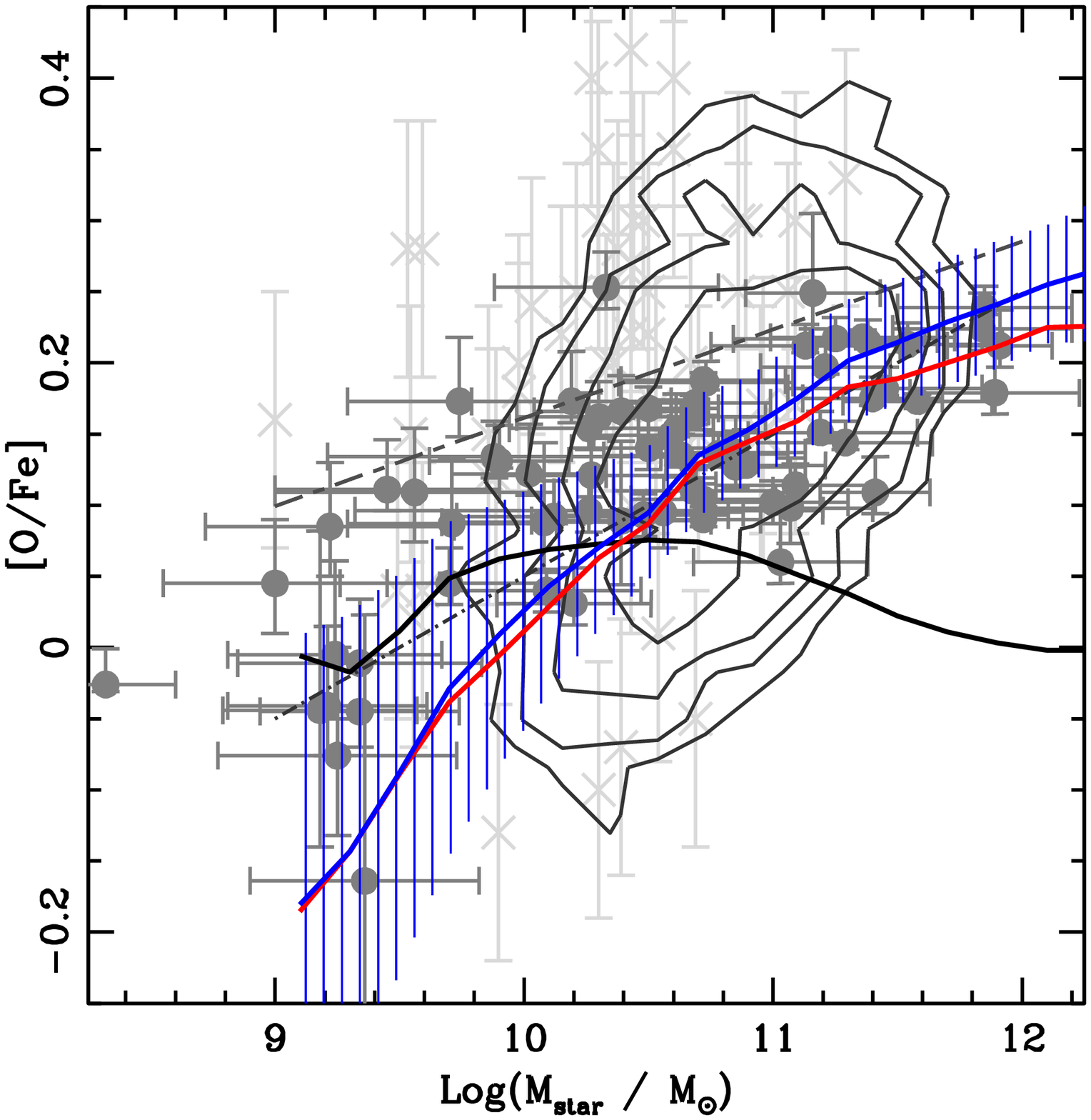} }
  \caption{[O/Fe] ratios as predicted by {\gaea} compared to observed
    [$\alpha$/Fe] ratios for local elliptical galaxies. Lines, colours
    and hatched area are as in Fig.~\ref{fig:mass-met} (only model
    galaxies with $B/T>0.7$ have been included in the sample). Grey
    symbols and contours represent data from \citet[][dark grey
      circles]{Arrigoni10}, \citet[][contours and dot-dashed
      line]{Thomas10}, \citet[][light grey crosses]{Spolaor10} and
    \citet[][long-short dashed
      line]{Johansson12}.}\label{fig:alpha_o_Fe}
\end{figure}

We then consider the [$\alpha$/Fe]-mass relation in early-type
galaxies (Fig.~\ref{fig:alpha_o_Fe}). We compare\footnote{In the
  following we will refer to [$\alpha$/Fe] ratios for the
  observational data and to the [O/Fe] for theoretical
  predictions. The rationale beyond this choice lies in the fact that,
  even if most of the observational estimates for [$\alpha$/Fe] are
  calibrated using Magnesium lines, Oxygen represents the most
  abundant among $\alpha$-elements.} the mean [O/Fe] versus stellar
mass relation in model galaxies with bulge-to-total ratios $B/T>0.7$
with the observational determinations for samples of local elliptical
galaxies from \citet{Arrigoni10}, \citet{Spolaor10}, \citet{Thomas10}
and \citet{Johansson12}. Abundance ratios are typically estimated
comparing spectral indexes (mainly Lick indexes) with predictions from
evolutionary spectral synthesis codes. The calibration of these codes
is critical for the correct recovery of element abundances. As an
example, in Fig.~\ref{fig:alpha_o_Fe} we show the estimates from
\citet[contours and dot-dashed line]{Thomas10} and \citet[long-short
  dashed line]{Johansson12}. These two studies consider the same
sample of 3360 $0.05 \le z \le 0.06$ early-type galaxies from the
MOSES (Morphologically Selected Early- types in SDSS) catalogue, but
they analysed the sample using different versions of the same stellar
population synthesis code \citep{Maraston05}. The difference in the
results is driven by the different calibration adopted for the
synthetic indexes (either based on globular clusters data or on
flux-calibrated stellar libraries, see \citealt{Thomas11a} for more
details), which impacts both the normalization and slope of the
[$\alpha$/Fe]-mass relation. The \citet{Thomas10} estimate is in good
agreement with the data in \citet{Arrigoni10}, who present a
re-analysis of the data in \citet{Trager00}, using a different stellar
population synthesis method \citep[i.e.][]{Trager08}. All available
measurements are obtained comparing data with synthetic spectra
derived under the universal IMF assumption; however, we do not expect
the systematic deviation due to the IGIMF to be larger than 0.1 dex
\citep{Recchi09}.

Stellar masses for individual sources in these datasets are derived
from measured velocity dispersions and formally represent dynamical
mass estimates. They are thus different from the proper $M_\star$
predicted by our models. Nonetheless, we do not expect this mismatch
to affect our main conclusion, as the stellar mass within one
effective radius is a good proxy of dynamical mass for an early type
galaxy \citep{Cappellari06}. The shape of the predicted
[$\alpha$/Fe]-mass relation is robust also if we consider
$M_\star^{\rm app}$. We remind that we did not require our model to
fit this relation as a part of the recalibration procedure. The
uncalibrated model (i.e. a model including the IGIMF, but using the
same parameters as HDLF16) shows the same positive trend of
[$\alpha$/Fe] with stellar mass. We also note that a similar
[O/Fe]-mass relation holds for the whole galaxy population, not only
for ellipticals galaxies.

While galaxies in HDLF16 are characterized by a flat distribution of
[O/Fe], our new version of {\gaea} nicely recovers the observed
trend. The hatched blue region marks the 1-$\sigma$ scatter around the
mean relation in the High-$\alpha_{\rm SF}$ run, and is representative of the
scatter in all our runs. Predictions from the IGIMF runs are in good
agreement with the \citet{Arrigoni10} and \citet{Thomas10} samples,
while they show a systematic offset from the more recent analysis by
\citet{Johansson12}. The \citet{Spolaor10} sample (based on data on
early-type galaxies in the Fornax and Virgo clusters) exhibit a larger
scatter with respect to both our model predictions and other
observational datasets.

In order to investigate the physical origin of the trends predicted by
our IGIMF runs, we consider two additional runs (for the
High-$\alpha_{\rm SF}$ parameter set), where we impose a Chabrier-IMF
either at $SFR<1M_\odot/yr$ or at $SFR>1M_\odot/yr$. In the former
run, we observe the same increase of [O/Fe] for $M_\star \gtrsim
10^{10} M_\odot$ as in the High-$\alpha_{\rm SF}$ case, while lower
mass galaxies show the same level of [O/Fe] as in the reference HDLF16
run. Viceversa, in the latter run, $M_\star \gtrsim 10^{10} M_\odot$
galaxies show the same flat distribution as in HDLF16, while lower
mass galaxies are characterized by a decrease in [O/Fe]. We then
conclude that the increase in $\alpha$-enhancement for massive
galaxies is due to the IMF being top-heavier than Chabrier in their
high-$\alpha_{\rm SF}$ events, and viceversa, the decrease of the
[$\alpha$/Fe] ratio in low-mass galaxies is due to these objects being
dominated by IMFs bottom-heavier than Chabrier. This result is
consistent with the similar analysis performed in G15, computing the
mean high-stellar mass slopes for the IMFs associated with galaxies of
different stellar mass. In their Fig.~5 they show that galaxies of
increasing stellar mass are characterized by mean slopes increasingly
top-heavier than low-mass counterparts. Although we use the same
functional form for the universal IMF in individual stellar clusters
(Eq.~\ref{eq:kroimf}), our IGIMF description features a four-sloped
shape (Sec.~\ref{sec:igimf}), that better describes the high-mass
trends. Therefore, we can not replicate the same analysis as in G15
with our models. Nonetheless, the comparison of our findings with G15
clearly show that the IGIMF approach provides consistent predictions
among different SAMs.

It is interesting to compare our findings with previous results from
\citet{Arrigoni10}, \citet{CaluraMenci11}, \citet{Yates13} and G15,
who claimed agreement between their model predictions and the observed
[$\alpha$/Fe]-stellar mass relation. These authors invoke different
mechanisms to explain the success of their models. The SAM discussed
in \citet{Arrigoni10} requires a mildly top-heavy IMF
(i.e. $\alpha_3=1.15$) and a fraction of binaries that explode as SNe
Ia of about 3 percent. \citet{CaluraMenci11} advocate that the
combination of ``fly-by'' harassment (that triggers starburst,
boosting the SFR at high-redshift) and AGN feedback (which efficiently
quench the same starbursts) considerably enhance the
[$\alpha$/Fe]-levels in massive galaxies. \citet{Yates13} reproduce a
positive slope for [$\alpha$/Fe]-stellar mass relation, consistent
with the \citet{Johansson12} estimate, using a universal
(Chabrier-like) IMF and assuming that the prompt component in the
SN-Ia delay-time distribution (DTD) is smaller than 50
percent. Finally, G15 implement an IGIMF model, which is rather
similar to our approach: the main difference with our work is that
they use the [$\alpha$/Fe]-mass relation for calibration, while in our
approach this is a genuine prediction of the model. G15 find an almost
flat [$\alpha$/Fe]-stellar mass relation for models adopting a
universal Salpeter-like IMF, while obtaining a noticeable steepening
of the relation for the IGIMF runs.

Predictions of our reference model are consistent with these results:
our standard model based on a universal IMF predicts slopes for the
[$\alpha$/Fe]-stellar mass relation which are too shallow. We also
test the effect of different DTDs, and find that for all DTDs
considered in \citet{DeLucia14} the HDLF16 model predicts a flat
relation (but with a different normalization). Given the fact that we
do not calibrate our IGIMF runs on the [O/Fe]-$M_\star$ relation, the
striking agreement of our predictions with observed data (even for the
overall normalization of the relation) is remarkable. Although this
success is not unique to our model, we note relevant differences with
respect to previous work. First of all, our IGIMF runs predict the
steepest slope for this relation. This effect is likely connected with
the assumed variation of $\beta$ for $SFR>1 M_\odot$ yr$^{-1}$
(Eq.~\ref{eq:beta}). We check this by computing the mean value of
$\beta$ for the integrated stellar populations corresponding to the
typical star formation histories at different mass scales (weighted
with the stellar mass formed at each epoch). We find a clear trend of
decreasing $<\beta>$ with increasing $M_\star$, with $<\beta>=2$ for
low-mass galaxies and $<\beta>\sim1.75$ for the highest mass galaxies
in our cosmological sample. This effect was not considered in G15,
who, however, tested the effect of a different (fixed) $\beta$,
finding that smaller values correspond to steeper relations. Moreover,
for stellar masses $M_\star<10^{10} M_\odot$ all models considered in
G15 tend to predict positive mean [O/Fe], with only a small fraction
of the model galaxies showing negative ratios, while these negative
ratios are found for a non-negligible fraction of both the
\citet{Arrigoni10} and \citet{Spolaor10} samples. The actual fraction
of galaxies below a given [O/Fe] threshold in {\gaea} depends on the
overall normalization of the [O/Fe]-$M_\star$ relation. We have
verified that in the IGIMF runs, the normalization of the relation
(and weakly its slope) depends on the assumed DTD. This is due to the
different amount of `prompt' Fe released to the ISM (see e.g. the
analysis in \citealt{DeLucia14}).

We stress again that the IGIMF approach does not represent a unique
solution for the [$\alpha$/Fe]-stellar mass conundrum in hierarchical
models of galaxy formation and evolution. We discussed some
alternative models earlier in this section. We also note that several
authors claim that AGN feedback can play a big role in setting high
[$\alpha$/Fe] ratio in massive galaxies via a sudden quenching of star
formation (\citealt{Pipino09}, \citealt{Segers16}, Hirschmann et
al. in preparation).

\subsection{Assembly histories and stellar mass estimates}
\begin{figure}
  \centerline{ \includegraphics[width=9cm]{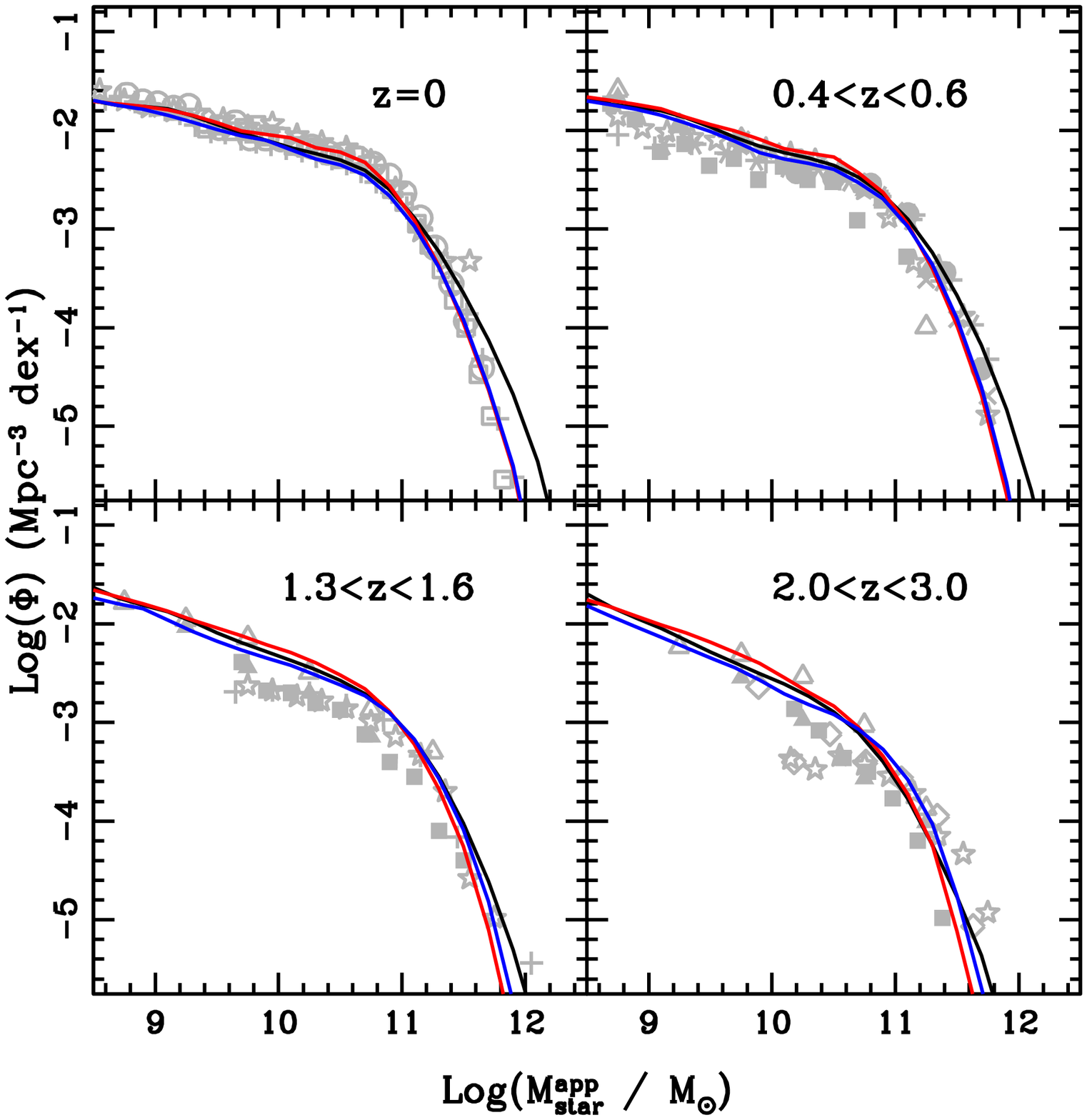} }
  \caption{Redshift evolution of the galaxy stellar mass function, as
    a function of the apparent $M_\star^{\rm app}$. Lines and colours
    are as in Fig.~\ref{fig:lfcal}; grey points refer to the
    compilation from \citet[][see references
      therein]{Fontanot09b}.}\label{fig:gsmf}
\end{figure}

In the previous section, we discussed the implications of the IGIMF
theory on the chemical enrichment properties of galaxies as predicted
by {\gaea}. In this section, we complement this investigation with
analysis of the effect on the distribution of stellar masses.

First of all we analyse the redshift evolution of the galaxy stellar
mass function. For all runs considered, when the {\it true} $M_\star$
is used, the predicted mass functions lie very close to those
predicted by HDLF16. On the other hand, when the photometrically
estimated, {\it apparent} $M_\star^{\rm app}$ is considered, the
high-mass end predictions from the IGIMF runs systematically deviate
from those of the HDLF16 model at $z<1$
(Fig.~\ref{fig:gsmf}). Overall, the IGIMF runs are able to reproduce
the evolution of the mass function as well as the reference HDLF16
runs, but the growth of the high mass-end of the mass function is
somewhat slowed down at $z\lesssim1$. Although the match between model
predictions and data is still not perfect, a reduced growth rate in
apparent $M_\star^{\rm app}$ for the most massive galaxies at
low-redshifts is an intriguing result \citep[see e.g.][]{Cimatti06,
  Monaco06}.

\begin{figure}
  \centerline{ \includegraphics[width=9cm]{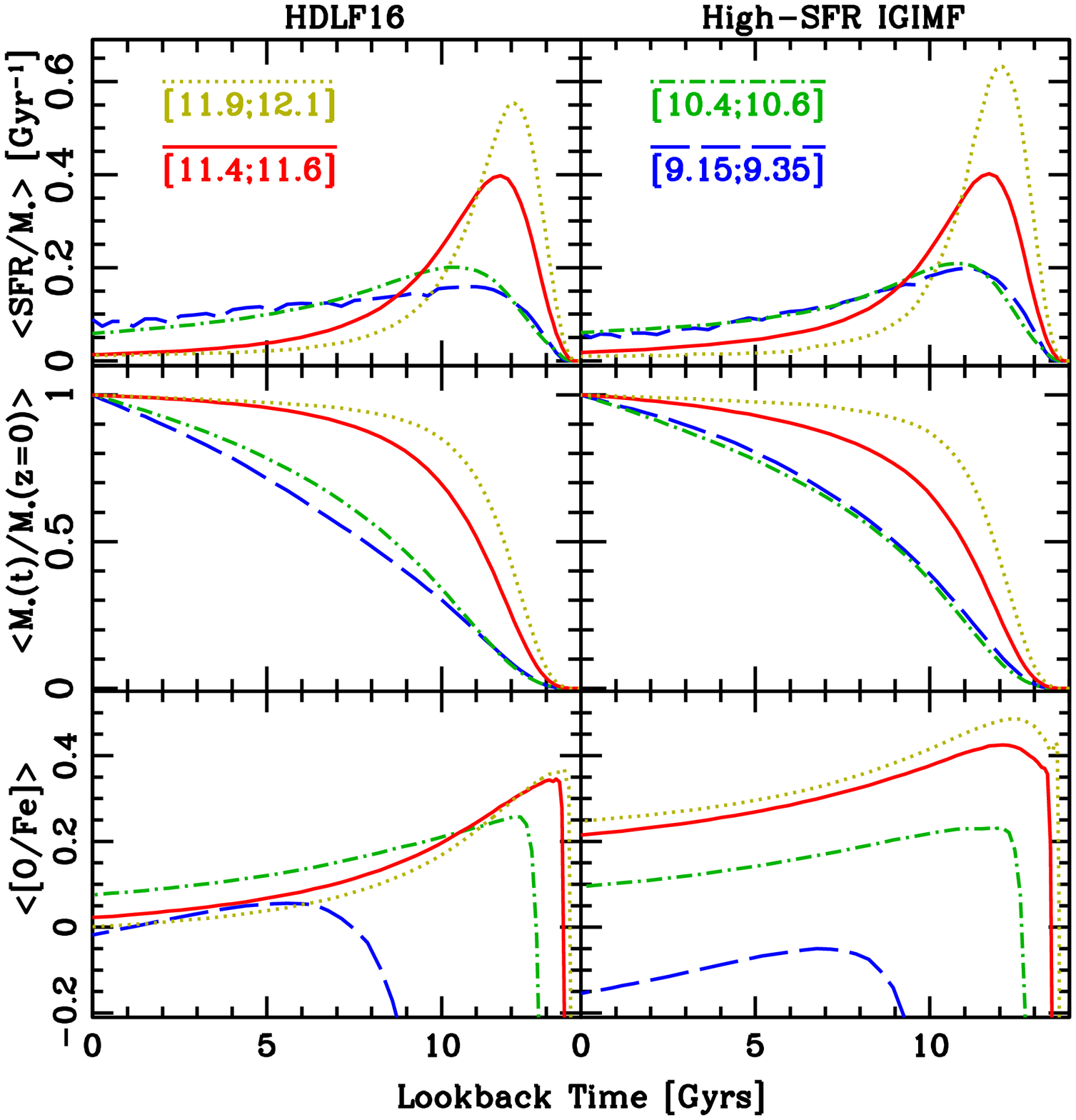} }
  \caption{Mean evolutionary histories for galaxies in different
    $Log(M/M_\odot)$ intervals (as indicate in the caption) in the
    HDLF16 (left column) and High-$\alpha_{\rm SF}$ (right column) runs. {\it Upper
      panels}: mean normalised star formation history; {\it middle
      panels}: cumulative mass assembly; {\it lower panels}: evolution
    of the [O/Fe] ratio.}\label{fig:sfh}
\end{figure}

In order to better understand the effect of a non universal IMF on
galaxy assembly, we contrast in Fig.~\ref{fig:sfh} the evolution of
key physical properties in the HDLF16 and High-$\alpha_{\rm SF}$ IGIMF
runs (similar results hold for the Low-$\alpha_{\rm SF}$ IGIMF run)
for model $B/T>0.7$ galaxies at four different $z=0$ mass bins
($M_\star \sim 10^{12}$, $10^{11.5}$, $10^{10.5}$ and $10^{9.25}
M_\odot$). In detail, we consider the normalised star formation
histories (upper panel), cumulative mass assembly (middle panel) and
the mean [O/Fe] (lower panel). For each model galaxy, the contribution
of all its progenitors has been included. Massive galaxies exhibit
star formation histories peaking at higher redshifts and a more rapid
assembly with respect to their low-mass counterparts; this is
consistent with results from \citet{DeLucia06} and predictions based
on the HDLF16 run. Therefore, the IGIMF scheme does not affect heavily
the overall star formation histories and mass assembly histories of
the different galaxy populations. The main difference is seen in the
lower panel: as the stronger SFRs associated with more massive
galaxies correspond to IGIMFs top-heavier than the universal IMF, this
results into a stronger $\alpha$-enhancement at earlier epochs with
respect to a model using a universal IMF. The later incorporation of
larger amounts of Fe produced by SNIa, produces a dilution of the mean
[O/Fe], but the tracks corresponding to the different mass scales
remain independent. On the other hand, in the HDLF16 run, the level of
initial $\alpha$-enhancement is reduced and late evolution tends to
wash differences out, leading to an average [O/Fe] ratio that does not
depend significantly on stellar mass. These results suggest that,
using the IGIMF, the so-called ``chemoarcheological downsizing''
naturally arise in a concordance cosmological model and that, under
the IGIMF assumption, [$\alpha$/Fe] ratios are good tracers of the
highest SFR events in galaxies of given mass, but they do not bear
much information on the overall star formation timescales. In fact, at
fixed $z=0$ stellar mass, the highest SFR events and the timescale of
star formation are degenerate quantities, but the former is the
dominant quantity to set the final level of $\alpha$-enhancement as a
function of stellar mass. This can be best appreciated in the
top-panels of Fig.~\ref{fig:sfh}: star formation timescales are
shorter for more massive galaxies both in the HDLF16 and the IGIMF
run, but only the latter realizations reach the required values of
[O/Fe].

We then directly study the effect of the IGIMF on the stellar mass
estimate. Independent analysis of early-type samples based either on
dynamical modeling \citep{Cappellari12} or on spectral synthesis
models \citep{ConroyvanDokkum12} suggest possible variations of the
overall shape of the IMF in these galaxies. In particular,
\citet{Cappellari12} compare integral-field maps of stellar kinematics
and optical imaging with dynamical models including both stellar and
DM components. \citet{ConroyvanDokkum12} consider a sample of compact
early-type galaxies (so that $\sigma$ is expected to be dominated by
the stellar component, at least within the effective radius) and
estimate mass-to-light ratios and stellar masses by fitting spectral
features sensitive to the stellar effective temperature and surface
gravity against stellar population synthesis models. Within this
framework, they model the high- and low-mass end of the IMF as free
parameters. They then compare the best-fit values for the physical
properties with those derived assuming a universal MW-like IMF, and
argue that the IMF in early-type galaxies becomes increasingly
``bottom-heavy'' (i.e. with a larger fraction of low mass stars with
respect to the universal IMF) with increasing velocity dispersion
($\sigma$) or galaxy stellar mass.
\begin{figure*}
  \centerline{ \includegraphics[width=9cm]{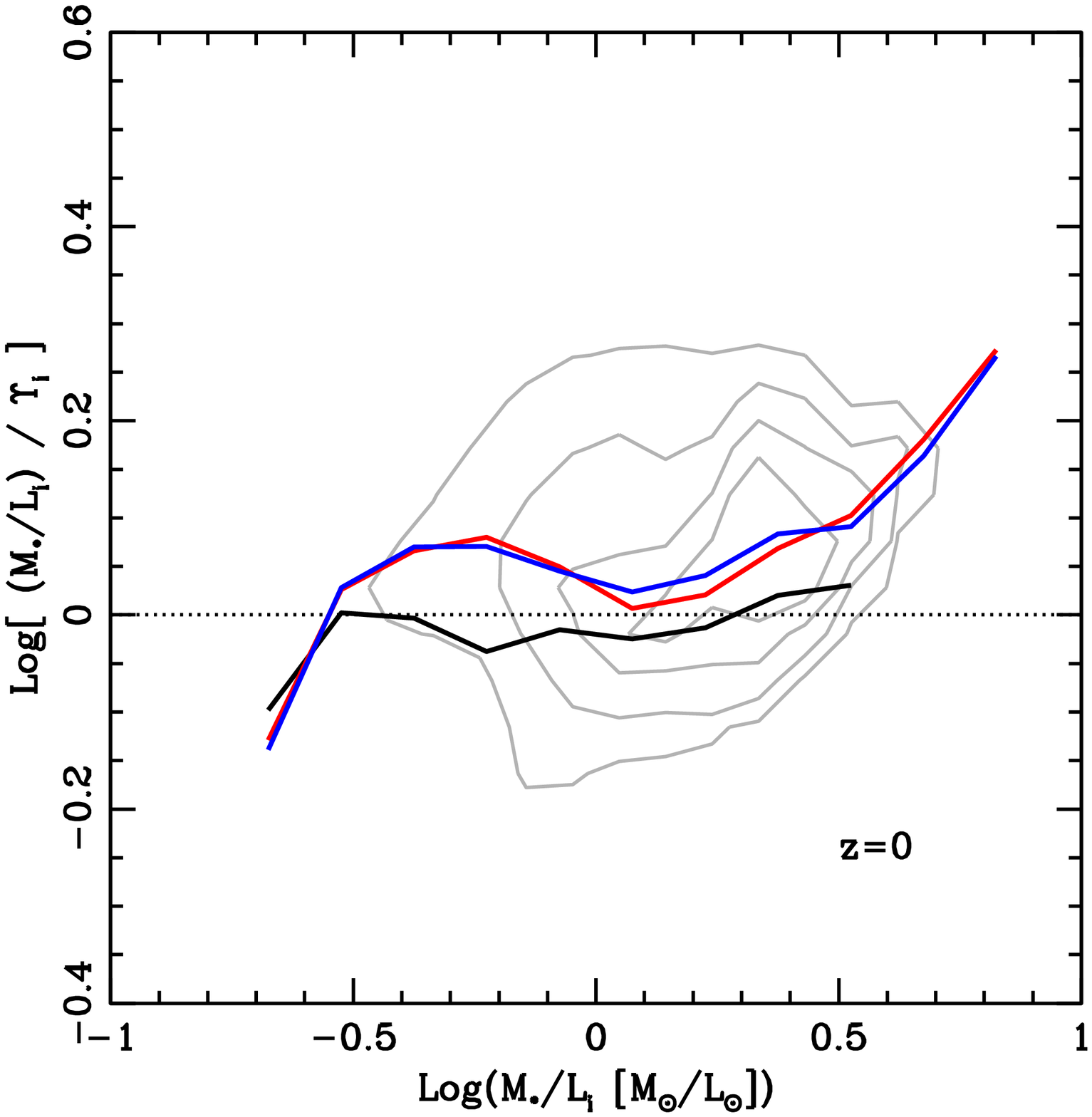} 
    \includegraphics[width=9cm]{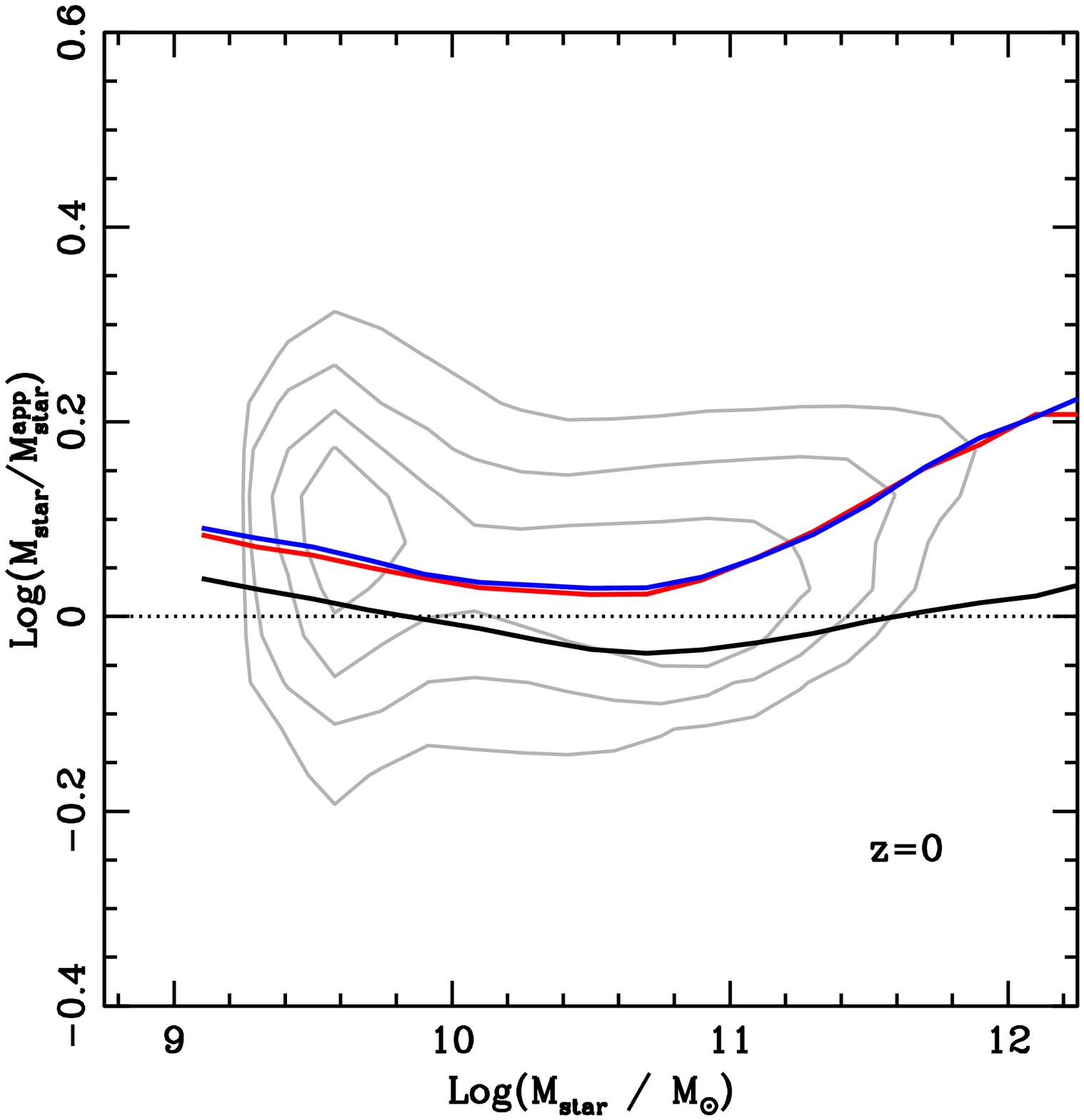} }
  \caption{Deviations from the assumption of a universal Chabrier-like
    IMF for stellar mass (right panel) and stellar mass-to-light ratio
    (left panel). In each panel, lines and colours are as in
    Fig.~\ref{fig:lfcal}; grey contours mark galaxy number densities
    levels (normalised to the maximum density) corresponding to 1, 10
    and 50 percent in the High-$\alpha_{\rm SF}$ IGIMF run.}\label{fig:demas}
\end{figure*}

In the left panel of Fig.~\ref{fig:demas} we show the ratio of the
proper stellar mass-to-light ratio in the $i$-band ($M_\star/L_i$) and
$\Upsilon_i$, derived using Eq.~\ref{eq:zibetti}, as a function of the
proper $M_\star/L_i$, as in the dynamical analysis of
\citet{Cappellari12}. The right panel shows the $M_\star /
M_\star^{\rm app}$ ratio, as a function of $M_\star$: this plot
roughly corresponds to the analogous figure in \citet{Conroy13}. In
both panels only bulge-dominated model galaxies (i.e. $B/T>0.7$) have
been considered. In both panels of Fig.~\ref{fig:demas} predictions
for the reference HDLF16 run are consistent with a flat relation,
while the IGIMF runs suggest a larger mass-to-light ratio (left panel)
and stellar mass (right panel) with respect to a \citet{Chabrier03}
IMF\footnote{We neglect here the small difference in normalization
  between the \citealt{Kroupa01} and the \citealt{Chabrier03} IMF used
  for the $M_\star^{\rm app}$ calibration} at increasing stellar mass
and/or mass-to-light ratio. Model predictions are therefore in good
qualitative agreement with the results of \citet{Cappellari12} and
\citet{ConroyvanDokkum12}. Moreover, we notice the relevant scatter in
the predicted relations, that may explain the results of
\citet{Smith15}, who analyse two gravitationally lensed massive
elliptical galaxies finding mass-to-light ratios consistent with a
universal IMF \citep[see also the recent results from][]{Leier15}.

It is worth stressing that the comparison of our model predictions
with data from \citet{Cappellari12} and \citet{ConroyvanDokkum12}.
can be only qualitative: a more quantitative comparison would require
an attempt to replicate both the same selection criteria adopted in
the observational studies and the same tracers for dynamical
properties (like $\sigma$). As a final note, we stress that
Fig.~\ref{fig:demas} differs from the similar plot shown in
\citet[][their Fig.~5]{Fontanot14}. In this work, the
photometrically-equivalent quantities are computed self-consistently
from the predicted magnitudes in the IGIMF realisation, while in
previous work the mass differences were computed comparing model
galaxies in different realizations (with or without a universal IMF)
on an object-by-object basis. The present approach provides a more
stringent constrain on the expected mass deviations with respect to a
local universal IMF.

\section{Conclusions}\label{sec:final}
This paper presents an updated version of the {\gaea} semi-analytic
model of galaxy formation and evolution, which includes the effects of
assuming that stars form following an IMF, whose shape depends on the
instantaneous SFR levels \citep[see e.g][]{Weidner13}. Coupled with
the detailed chemical enrichment model introduced in \citet{DeLucia14}
and with the feedback scheme presented in \citet{Hirschmann16}, this
version allows us to study the impact of this hypothesis on the galaxy
mass assembly and its imprint on the chemical abundances of both stars
and cold gas in galaxies.

The different amount of stars locked in a low-mass and long-living
population affects significantly the physical properties of model
galaxies, and forces a recalibration of the key free parameters
describing star formation and feedback. We choose to recalibrate our
{\gaea} version requiring it to reproduce the redshift evolution of
the $K$ and $V$ band luminosity functions. We then derive
photometrically-equivalent stellar masses using an empirical relation
between mass-to-light ratios and colours, calibrated using a large
library of synthetic spectra built using a universal
\citet{Chabrier03} IMF.

We show that this new model predicts local scalings of the
luminosity-weighted age, stellar metallicity and cold gas metallicity
with stellar mass in close agreement with the results of HDLF16. The
main difference with previous versions of {\gaea} lies in the
$\alpha$-enhancement of bulge-dominated galaxies: while HDLF16 predict
a flat relation with stellar mass, our IGIMF-based model correctly
reproduces the measured increase of [$\alpha$/Fe] ratios as a function
of stellar mass. These results confirm early findings by G15 and show
that the impact of the IGIMF approach is robust and independent of the
details of the semi-analytic model in which it has been
implemented. We also study the relation between the proper stellar
masses predicted by our best run and the apparent stellar masses
derived from synthetic photometry in the IGIMF runs, assuming a
universal IMF. We show that, for high-mass galaxies, the $M_\star /
M_\star^{\rm app}$ ratio is typically positive, with a relevant
scatter. Similar conclusions hold for the corresponding ratio between
the proper and apparent mass-to-light ratios. These predictions are in
qualitative agreement with data from \citet{Cappellari12} and
\citet{ConroyvanDokkum12}. These groups find in their data an excess
of mass-to-light ratio and stellar mass (respectively) with respect to
what expected using a universal IMF and interpret this discrepancy as
an evidence in favour of a typical IMF ``bottom-heavier'' than the
Chabrier or Kroupa IMF. In the framework of the IGIMF runs, the
discrepancy between $M_\star$ and $M_\star^{\rm app}$ is {\it not} due
to an intrinsic ``bottom-heavier'' IMF in massive galaxies. In fact,
our massive model galaxies are characterized by an effective IMF
(i.e. the mean slopes computed over the typical star formation
histories in Fig.~\ref{fig:sfh}) with high-mass slopes smaller than
2.35 (i.e. by a ``top-heavy'' IMF). Our conclusions are therefore in
contrast with those by \citet{Cappellari12}
and~\citet{ConroyvanDokkum12}, which we interpret as due to the
mismatch between proper mass-to-light ratios and those derived from
synthetic photometry, under the assumption of a universal IMF. The
disagreement with \citet{ConroyvanDokkum12} is particularly
interesting, since these authors use spectral features sensitive to
the ratio between low-mass stars and giants, but we cannot explicitly
test these observables and quantify the level of disagreement, within
our current model predictions.

We test the robustness of our results against a change in the
modelling of mass and energy transfer between galaxy components
(bulge, disc, halo) in galaxy mergers as proposed in
\citet{Kannan15}. In \citet{Fontanot15b}, we showed that these new
prescriptions have a relevant impact on the distribution of galaxies
in the different morphological types. We then run an additional
realisation switching on \citet{Kannan15} recipes and using the same
parameters as in our IGIMF run (i.e. we did not attempt to recalibrate
this model). All predictions shown in this paper are robust against
this change in the merger modelling. The main difference we see is a
slight increase of the [$\alpha$/Fe]-enhancement at both the low-mass
and high-mass end, which brings model predictions in better agreement
with the linear fit of \citet{Thomas10}. These changes are not driven
by a different star formation history in this run, but from its
different sample of ``elliptical'' ($B/T>0.7$) galaxies.

In this paper, we show for the first time a model which reproduces,
{\it at the same time}, the evolution of the stellar mass function and
the abundance patterns of elliptical galaxies, and explains the
observed peculiar dynamical properties of local early type
galaxies. Our model is, however, not without problems. For example,
model galaxies below the knee of the mass function host stellar
populations that are still too old (irrespective of the feedback
scheme and treatment of satellite galaxies), indicating a disagreement
between predicted and observationally estimated star formation
histories at this mass scale.

As mentioned above, a varying IMF does not represent a unique solution
to reproduce the observed trend of [$\alpha$/Fe] ratios in early-type
galaxies: alternative solutions cannot be excluded and will be tested
in future work. Among these, metal enriched winds represent an
interesting option in the framework of strong ejective feedback models
like that adopted in {\gaea} \citep[see e.g.][]{Yates13}. Robust
constraints for different schemes can be obtained from the metal
enrichment of the intergalactic medium, typically traced using quasar
spectra (see e.g. \citealt{DOdorico16} and references herein). On
the other hand, dynamical studies provide the strongest indication in
favour of a varying IMF hypothesis. While a more quantitative
comparison of our model predictions with available data is beyond the
aims of the present work, a better characterisation of the selection
effects at play and a detailed modeling of physical quantities like
$\sigma$ are clearly required. Ongoing integral field spectroscopy
observations of large samples of nearby galaxies
(e.g. MaNGA\footnote{Mapping Nearby Galaxies at Apache Point
  Observatory}, \citealt{Bundy15}, or the SAMI\footnote{Sydney
  Australian Astronomical Observatory Multi-object Integral-Field
  Spectrograph} Galaxy survey, \citealt{Bryant15}) will provide
statistical support to the claimed excess of low-mass stars in massive
galaxies.

\section*{Acknowledgements}

We thank P. Kroupa and J. Pflamm-Altenburg for enlightening
discussions on the details of the IGIMF model and R. Yates for
stimulating discussions. We thank the anonymous referee for useful
comments that helped us improving the clarity of the manuscript. FF
acknowledges financial support from the grant PRIN MIUR 2012 ``The
Intergalactic Medium as a probe of the growth of cosmic
structures''. GDL acknowledges financial support from the MERAC
foundation. FF and GDL also acknowledge from the grants PRIN INAF 2014
``Glittering kaleidoscopes in the sky: the multifaceted nature and
role of Galaxy Clusters.'' MH acknowledges financial support from the
European Research Council via an Advanced Grant under grant agreement
no. 321323 (NEOGAL). GB acknowledges support for this work from the
National Autonomous University of M\'exico (UNAM), through grant
PAPIIT IG100115. SZ has been supported by the EU Marie Curie Career
Integration Grant ``SteMaGE'' Nr. PCIG12-GA-2012-326466 (Call
Identifier: FP7-PEOPLE-2012 CIG)

\bibliographystyle{mn2e}
\bibliography{fontanot}

\end{document}


%% file: imfvar_rev2.bbl
\begin{thebibliography}{}

\bibitem[\protect\citeauthoryear{{Arrigoni}, {Trager}, {Somerville} \&
  {Gibson}}{{Arrigoni} et~al.}{2010}]{Arrigoni10}
{Arrigoni} M.,  {Trager} S.~C.,  {Somerville} R.~S.,    {Gibson} B.~K.,  2010,
  \mnras, 402, 173

\bibitem[\protect\citeauthoryear{{Blanton}, {Lupton}, {Schlegel}, {Strauss},
  {Brinkmann}, {Fukugita} \& {Loveday}}{{Blanton} et~al.}{2005}]{Blanton04}
{Blanton} M.~R.,  {Lupton} R.~H.,  {Schlegel} D.~J.,  {Strauss} M.~A.,
  {Brinkmann} J.,  {Fukugita} M.,    {Loveday} J.,  2005, \apj, 631, 208

\bibitem[\protect\citeauthoryear{{Boselli}, {Cortese}, {Boquien}, {Boissier},
  {Catinella}, {Lagos} \& {Saintonge}}{{Boselli} et~al.}{2014}]{Boselli14}
{Boselli} A.,  {Cortese} L.,  {Boquien} M.,  {Boissier} S.,  {Catinella} B.,
  {Lagos} C.,    {Saintonge} A.,  2014, \aap, 564, A66

\bibitem[\protect\citeauthoryear{{Bruzual} \& {Charlot}}{{Bruzual} \&
  {Charlot}}{2003}]{Bruzual03}
{Bruzual} G.,  {Charlot} S.,  2003, \mnras, 344, 1000

\bibitem[\protect\citeauthoryear{{Bryant} \& {et al.}}{{Bryant} \& {et
  al.}}{2015}]{Bryant15}
{Bryant} J.~J.,  {et al.} 2015, \mnras, 447, 2857

\bibitem[\protect\citeauthoryear{{Bundy} \& {et al.}}{{Bundy} \& {et
  al.}}{2015}]{Bundy15}
{Bundy} K.,  {et al.} 2015, \apj, 798, 7

\bibitem[\protect\citeauthoryear{{Calura} \& {Menci}}{{Calura} \&
  {Menci}}{2011}]{CaluraMenci11}
{Calura} F.,  {Menci} N.,  2011, \mnras, 413, L1

\bibitem[\protect\citeauthoryear{{Cappellari}, {Bacon}, {Bureau}, {Damen},
  {Davies}, {de Zeeuw}, {Emsellem}, {Falc{\'o}n-Barroso}, {Krajnovi{\'c}},
  {Kuntschner}, {McDermid}, {Peletier}, {Sarzi}, {van den Bosch} \& {van de
  Ven}}{{Cappellari} et~al.}{2006}]{Cappellari06}
{Cappellari} M.,  {Bacon} R.,  {Bureau} M.,  {Damen} M.~C.,  {Davies} R.~L.,
  {de Zeeuw} P.~T.,  {Emsellem} E.,  {Falc{\'o}n-Barroso} J.,  {Krajnovi{\'c}}
  D.,  {Kuntschner} H.,  {McDermid} R.~M.,  {Peletier} R.~F.,  {Sarzi} M.,
  {van den Bosch} R.~C.~E.,    {van de Ven} G.,  2006, \mnras, 366, 1126

\bibitem[\protect\citeauthoryear{{Cappellari}, {McDermid}, {Alatalo}, {Blitz},
  {Bois}, {Bournaud}, {Bureau}, {Crocker} \& et al.}{{Cappellari}
  et~al.}{2012}]{Cappellari12}
{Cappellari} M.,  {McDermid} R.~M.,  {Alatalo} K.,  {Blitz} L.,  {Bois} M.,
  {Bournaud} F.,  {Bureau} M.,  {Crocker} A.~F.,    et al. 2012, \nat, 484, 485

\bibitem[\protect\citeauthoryear{{Chabrier}}{{Chabrier}}{2003}]{Chabrier03}
{Chabrier} G.,  2003, \apjl, 586, L133

\bibitem[\protect\citeauthoryear{{Chieffi} \& {Limongi}}{{Chieffi} \&
  {Limongi}}{2002}]{ChieffiLimongi02}
{Chieffi} A.,  {Limongi} M.,  2002, \apj, 577, 281

\bibitem[\protect\citeauthoryear{{Cimatti}, {Daddi} \& {Renzini}}{{Cimatti}
  et~al.}{2006}]{Cimatti06}
{Cimatti} A.,  {Daddi} E.,    {Renzini} A.,  2006, \aap, 453, L29

\bibitem[\protect\citeauthoryear{{Cirasuolo}, {McLure}, {Dunlop}, {Almaini},
  {Foucaud} \& {Simpson}}{{Cirasuolo} et~al.}{2010}]{Cirasuolo10}
{Cirasuolo} M.,  {McLure} R.~J.,  {Dunlop} J.~S.,  {Almaini} O.,  {Foucaud} S.,
     {Simpson} C.,  2010, \mnras, 401, 1166

\bibitem[\protect\citeauthoryear{{Cole}, {Norberg}, {Baugh}, {Frenk},
  {Bland-Hawthorn}, {Bridges}, {Cannon} \& {Colless}}{{Cole}
  et~al.}{2001}]{Cole01}
{Cole} S.,  {Norberg} P.,  {Baugh} C.~M.,  {Frenk} C.~S.,  {Bland-Hawthorn} J.,
   {Bridges} T.,  {Cannon} R.,    {Colless} M. e.~a.,  2001, \mnras, 326, 255

\bibitem[\protect\citeauthoryear{{Conroy}, {Dutton}, {Graves}, {Mendel} \& {van
  Dokkum}}{{Conroy} et~al.}{2013}]{Conroy13}
{Conroy} C.,  {Dutton} A.~A.,  {Graves} G.~J.,  {Mendel} J.~T.,    {van Dokkum}
  P.~G.,  2013, \apjl, 776, L26

\bibitem[\protect\citeauthoryear{{Conroy} \& {van Dokkum}}{{Conroy} \& {van
  Dokkum}}{2012}]{ConroyvanDokkum12}
{Conroy} C.,  {van Dokkum} P.~G.,  2012, \apj, 760, 71

\bibitem[\protect\citeauthoryear{{Cora}}{{Cora}}{2006}]{Cora06}
{Cora} S.~A.,  2006, \mnras, 368, 1540

\bibitem[\protect\citeauthoryear{{De Lucia} \& {Blaizot}}{{De Lucia} \&
  {Blaizot}}{2007}]{DeLuciaBlaizot07}
{De Lucia} G.,  {Blaizot} J.,  2007, \mnras, 375, 2

\bibitem[\protect\citeauthoryear{{De Lucia} \& {Helmi}}{{De Lucia} \&
  {Helmi}}{2008}]{DeLuciaHelmi08}
{De Lucia} G.,  {Helmi} A.,  2008, \mnras, 391, 14

\bibitem[\protect\citeauthoryear{{De Lucia}, {Kauffmann} \& {White}}{{De Lucia}
  et~al.}{2004}]{DeLucia04b}
{De Lucia} G.,  {Kauffmann} G.,    {White} S.~D.~M.,  2004, \mnras, 349, 1101

\bibitem[\protect\citeauthoryear{{De Lucia}, {Springel}, {White}, {Croton} \&
  {Kauffmann}}{{De Lucia} et~al.}{2006}]{DeLucia06}
{De Lucia} G.,  {Springel} V.,  {White} S.~D.~M.,  {Croton} D.,    {Kauffmann}
  G.,  2006, \mnras, 366, 499

\bibitem[\protect\citeauthoryear{{De Lucia}, {Tornatore}, {Frenk}, {Helmi},
  {Navarro} \& {White}}{{De Lucia} et~al.}{2014}]{DeLucia14}
{De Lucia} G.,  {Tornatore} L.,  {Frenk} C.~S.,  {Helmi} A.,  {Navarro} J.~F.,
    {White} S.~D.~M.,  2014, \mnras, 445, 970

\bibitem[\protect\citeauthoryear{{D'Odorico}, {Cristiani}, {Pomante},
  {Carswell}, {Viel}, {Barai}, {Becker}, {Calura}, {Cupani}, {Fontanot},
  {Haehnelt}, {Kim}, {Miralda-Escud{\'e}}, {Rorai}, {Tescari} \&
  {Vanzella}}{{D'Odorico} et~al.}{2016}]{DOdorico16}
{D'Odorico} V.,  {Cristiani} S.,  {Pomante} E.,  {Carswell} R.~F.,  {Viel} M.,
  {Barai} P.,  {Becker} G.~D.,  {Calura} F.,  {Cupani} G.,  {Fontanot} F.,
  {Haehnelt} M.~G.,  {Kim} T.-S.,  {Miralda-Escud{\'e}} J.,  {Rorai} A.,
  {Tescari} E.,    {Vanzella} E.,  2016, \mnras

\bibitem[\protect\citeauthoryear{{Erb}, {Shapley}, {Pettini}, {Steidel},
  {Reddy} \& {Adelberger}}{{Erb} et~al.}{2006}]{Erb06}
{Erb} D.~K.,  {Shapley} A.~E.,  {Pettini} M.,  {Steidel} C.~C.,  {Reddy} N.~A.,
     {Adelberger} K.~L.,  2006, \apj, 644, 813

\bibitem[\protect\citeauthoryear{{Ferreras}, {La Barbera}, {de la Rosa},
  {Vazdekis}, {de Carvalho}, {Falc{\'o}n-Barroso} \& {Ricciardelli}}{{Ferreras}
  et~al.}{2013}]{Ferreras13}
{Ferreras} I.,  {La Barbera} F.,  {de la Rosa} I.~G.,  {Vazdekis} A.,  {de
  Carvalho} R.~R.,  {Falc{\'o}n-Barroso} J.,    {Ricciardelli} E.,  2013,
  \mnras, 429, L15

\bibitem[\protect\citeauthoryear{{Fontanot}}{{Fontanot}}{2014}]{Fontanot14}
{Fontanot} F.,  2014, \mnras, 442, 3138

\bibitem[\protect\citeauthoryear{{Fontanot}, {De Lucia}, {Monaco}, {Somerville}
  \& {Santini}}{{Fontanot} et~al.}{2009}]{Fontanot09b}
{Fontanot} F.,  {De Lucia} G.,  {Monaco} P.,  {Somerville} R.~S.,    {Santini}
  P.,  2009, \mnras, 397, 1776

\bibitem[\protect\citeauthoryear{{Fontanot}, {Macci{\`o}}, {Hirschmann}, {De
  Lucia}, {Kannan}, {Somerville} \& {Wilman}}{{Fontanot}
  et~al.}{2015}]{Fontanot15b}
{Fontanot} F.,  {Macci{\`o}} A.~V.,  {Hirschmann} M.,  {De Lucia} G.,  {Kannan}
  R.,  {Somerville} R.~S.,    {Wilman} D.,  2015, \mnras, 451, 2968

\bibitem[\protect\citeauthoryear{{Gallazzi}, {Charlot}, {Brinchmann}, {White}
  \& {Tremonti}}{{Gallazzi} et~al.}{2005}]{Gallazzi05}
{Gallazzi} A.,  {Charlot} S.,  {Brinchmann} J.,  {White} S.~D.~M.,
  {Tremonti} C.~A.,  2005, \mnras, 362, 41

\bibitem[\protect\citeauthoryear{{Gargiulo}, {Cora}, {Padilla}, {Mu{\~n}oz
  Arancibia}, {Ruiz}, {Orsi}, {Tecce}, {Weidner} \& {Bruzual}}{{Gargiulo}
  et~al.}{2015}]{Gargiulo14}
{Gargiulo} I.~D.,  {Cora} S.~A.,  {Padilla} N.~D.,  {Mu{\~n}oz Arancibia}
  A.~M.,  {Ruiz} A.~N.,  {Orsi} A.~A.,  {Tecce} T.~E.,  {Weidner} C.,
  {Bruzual} G.,  2015, \mnras, 446, 3820

\bibitem[\protect\citeauthoryear{{Gunawardhana}, {Hopkins}, {Sharp}, {Brough},
  {Taylor}, {Bland-Hawthorn}, {Maraston}, {Tuffs} \& et al.}{{Gunawardhana}
  et~al.}{2011}]{Gunawardhana11}
{Gunawardhana} M.~L.~P.,  {Hopkins} A.~M.,  {Sharp} R.~G.,  {Brough} S.,
  {Taylor} E.,  {Bland-Hawthorn} J.,  {Maraston} C.,  {Tuffs} R.~J.,    et al.
  2011, \mnras, 415, 1647

\bibitem[\protect\citeauthoryear{{Guo}, {White}, {Angulo}, {Henriques},
  {Lemson}, {Boylan-Kolchin}, {Thomas} \& {Short}}{{Guo} et~al.}{2013}]{Guo13}
{Guo} Q.,  {White} S.,  {Angulo} R.~E.,  {Henriques} B.,  {Lemson} G.,
  {Boylan-Kolchin} M.,  {Thomas} P.,    {Short} C.,  2013, \mnras, 428, 1351

\bibitem[\protect\citeauthoryear{{Guo}, {White}, {Boylan-Kolchin}, {De Lucia},
  {Kauffmann}, {Lemson}, {Li}, {Springel} \& {Weinmann}}{{Guo}
  et~al.}{2011}]{Guo11}
{Guo} Q.,  {White} S.,  {Boylan-Kolchin} M.,  {De Lucia} G.,  {Kauffmann} G.,
  {Lemson} G.,  {Li} C.,  {Springel} V.,    {Weinmann} S.,  2011, \mnras, 413,
  101

\bibitem[\protect\citeauthoryear{{Hennebelle} \& {Chabrier}}{{Hennebelle} \&
  {Chabrier}}{2008}]{HennebelleChabrier08}
{Hennebelle} P.,  {Chabrier} G.,  2008, \apj, 684, 395

\bibitem[\protect\citeauthoryear{{Henriques}, {White}, {Thomas}, {Angulo},
  {Guo}, {Lemson} \& {Springel}}{{Henriques} et~al.}{2013}]{Henriques13}
{Henriques} B.~M.~B.,  {White} S.~D.~M.,  {Thomas} P.~A.,  {Angulo} R.~E.,
  {Guo} Q.,  {Lemson} G.,    {Springel} V.,  2013, \mnras, 431, 3373

\bibitem[\protect\citeauthoryear{{Hirschmann}, {De Lucia} \&
  {Fontanot}}{{Hirschmann} et~al.}{2015}]{Hirschmann16}
{Hirschmann} M.,  {De Lucia} G.,    {Fontanot} F.,  2015, ArXiv e-prints
  (arXiv:1512.04531)

\bibitem[\protect\citeauthoryear{{Hopkins}}{{Hopkins}}{2012}]{Hopkins12}
{Hopkins} P.~F.,  2012, \mnras, 423, 2037

\bibitem[\protect\citeauthoryear{{Hopkins}, {Kere{\v s}}, {O{\~n}orbe},
  {Faucher-Gigu{\`e}re}, {Quataert}, {Murray} \& {Bullock}}{{Hopkins}
  et~al.}{2014}]{Hopkins14}
{Hopkins} P.~F.,  {Kere{\v s}} D.,  {O{\~n}orbe} J.,  {Faucher-Gigu{\`e}re}
  C.-A.,  {Quataert} E.,  {Murray} N.,    {Bullock} J.~S.,  2014, \mnras, 445,
  581

\bibitem[\protect\citeauthoryear{{Hoversten} \& {Glazebrook}}{{Hoversten} \&
  {Glazebrook}}{2008}]{HoverstenGlazebrook08}
{Hoversten} E.~A.,  {Glazebrook} K.,  2008, \apj, 675, 163

\bibitem[\protect\citeauthoryear{{Johansson}, {Thomas} \&
  {Maraston}}{{Johansson} et~al.}{2012}]{Johansson12}
{Johansson} J.,  {Thomas} D.,    {Maraston} C.,  2012, \mnras, 421, 1908

\bibitem[\protect\citeauthoryear{{Kannan}, {Macci{\`o}}, {Fontanot}, {Moster},
  {Karman} \& {Somerville}}{{Kannan} et~al.}{2015}]{Kannan15}
{Kannan} R.,  {Macci{\`o}} A.~V.,  {Fontanot} F.,  {Moster} B.~P.,  {Karman}
  W.,    {Somerville} R.~S.,  2015, \mnras, 452, 4347

\bibitem[\protect\citeauthoryear{{Karakas}}{{Karakas}}{2010}]{Karakas10}
{Karakas} A.~I.,  2010, \mnras, 403, 1413

\bibitem[\protect\citeauthoryear{{Kewley} \& {Ellison}}{{Kewley} \&
  {Ellison}}{2008}]{KewleyEllison08}
{Kewley} L.~J.,  {Ellison} S.~L.,  2008, \apj, 681, 1183

\bibitem[\protect\citeauthoryear{{Klessen}, {Ballesteros-Paredes},
  {V{\'a}zquez-Semadeni} \& {Dur{\'a}n-Rojas}}{{Klessen}
  et~al.}{2005}]{Klessen05}
{Klessen} R.~S.,  {Ballesteros-Paredes} J.,  {V{\'a}zquez-Semadeni} E.,
  {Dur{\'a}n-Rojas} C.,  2005, \apj, 620, 786

\bibitem[\protect\citeauthoryear{{Klessen}, {Spaans} \& {Jappsen}}{{Klessen}
  et~al.}{2007}]{Klessen07}
{Klessen} R.~S.,  {Spaans} M.,    {Jappsen} A.-K.,  2007, \mnras, 374, L29

\bibitem[\protect\citeauthoryear{{Kochanek}, {Pahre}, {Falco}, {Huchra},
  {Mader}, {Jarrett}, {Chester}, {Cutri} \& {Schneider}}{{Kochanek}
  et~al.}{2001}]{Kochanek01}
{Kochanek} C.~S.,  {Pahre} M.~A.,  {Falco} E.~E.,  {Huchra} J.~P.,  {Mader} J.,
   {Jarrett} T.~H.,  {Chester} T.,  {Cutri} R.,    {Schneider} S.~E.,  2001,
  \apj, 560, 566

\bibitem[\protect\citeauthoryear{{Kroupa}}{{Kroupa}}{2001}]{Kroupa01}
{Kroupa} P.,  2001, \mnras, 322, 231

\bibitem[\protect\citeauthoryear{{Kroupa} \& {Bouvier}}{{Kroupa} \&
  {Bouvier}}{2003}]{KroupaBouvier03}
{Kroupa} P.,  {Bouvier} J.,  2003, \mnras, 346, 369

\bibitem[\protect\citeauthoryear{{Kroupa}, {Weidner}, {Pflamm-Altenburg},
  {Thies}, {Dabringhausen}, {Marks} \& {Maschberger}}{{Kroupa}
  et~al.}{2013}]{Kroupa13}
{Kroupa} P.,  {Weidner} C.,  {Pflamm-Altenburg} J.,  {Thies} I.,
  {Dabringhausen} J.,  {Marks} M.,    {Maschberger} T.,  2013, Planets, Stars
  and Stellar Systems.~Volume 5: Galactic Structure and Stellar Populations, 5,
  115

\bibitem[\protect\citeauthoryear{{Krumholz}}{{Krumholz}}{2014}]{Krumholz14}
{Krumholz} M.~R.,  2014, ArXiv e-prints (arXiv:1402.0867)

\bibitem[\protect\citeauthoryear{{Lada} \& {Lada}}{{Lada} \&
  {Lada}}{2003}]{LadaLada03}
{Lada} C.~J.,  {Lada} E.~A.,  2003, \araa, 41, 57

\bibitem[\protect\citeauthoryear{{Leier}, {Ferreras}, {Saha}, {Charlot},
  {Bruzual} \& {La Barbera}}{{Leier} et~al.}{2015}]{Leier15}
{Leier} D.,  {Ferreras} I.,  {Saha} P.,  {Charlot} S.,  {Bruzual} G.,    {La
  Barbera} F.,  2015, ArXiv e-prints

\bibitem[\protect\citeauthoryear{{Li}, {De Lucia} \& {Helmi}}{{Li}
  et~al.}{2010}]{Li10}
{Li} Y.-S.,  {De Lucia} G.,    {Helmi} A.,  2010, \mnras, 401, 2036

\bibitem[\protect\citeauthoryear{{Loveday}, {Norberg}, {Baldry}, {Driver},
  {Hopkins}, {Peacock}, {Bamford}, {Liske} \& et al.}{{Loveday}
  et~al.}{2012}]{Loveday12}
{Loveday} J.,  {Norberg} P.,  {Baldry} I.~K.,  {Driver} S.~P.,  {Hopkins}
  A.~M.,  {Peacock} J.~A.,  {Bamford} S.~P.,  {Liske} J.,    et al. 2012,
  \mnras, 420, 1239

\bibitem[\protect\citeauthoryear{{Maraston}}{{Maraston}}{2005}]{Maraston05}
{Maraston} C.,  2005, \mnras, 362, 799

\bibitem[\protect\citeauthoryear{{Marchesini}, {Stefanon}, {Brammer} \&
  {Whitaker}}{{Marchesini} et~al.}{2012}]{Marchesini12}
{Marchesini} D.,  {Stefanon} M.,  {Brammer} G.~B.,    {Whitaker} K.~E.,  2012,
  \apj, 748, 126

\bibitem[\protect\citeauthoryear{{Marigo}, {Girardi}, {Bressan}, {Groenewegen},
  {Silva} \& {Granato}}{{Marigo} et~al.}{2008}]{Marigo08}
{Marigo} P.,  {Girardi} L.,  {Bressan} A.,  {Groenewegen} M.~A.~T.,  {Silva}
  L.,    {Granato} G.~L.,  2008, \aap, 482, 883

\bibitem[\protect\citeauthoryear{{Marks} \& {Kroupa}}{{Marks} \&
  {Kroupa}}{2012}]{MarksKroupa12}
{Marks} M.,  {Kroupa} P.,  2012, \aap, 543, A8

\bibitem[\protect\citeauthoryear{{Marks}, {Kroupa}, {Dabringhausen} \&
  {Pawlowski}}{{Marks} et~al.}{2012}]{Marks12}
{Marks} M.,  {Kroupa} P.,  {Dabringhausen} J.,    {Pawlowski} M.~S.,  2012,
  \mnras, 422, 2246

\bibitem[\protect\citeauthoryear{{Matteucci}}{{Matteucci}}{1994}]{Matteucci94}
{Matteucci} F.,  1994, \aap, 288, 57

\bibitem[\protect\citeauthoryear{{Matteucci} \& {Recchi}}{{Matteucci} \&
  {Recchi}}{2001}]{MatteucciRecchi01}
{Matteucci} F.,  {Recchi} S.,  2001, \apj, 558, 351

\bibitem[\protect\citeauthoryear{{McWilliam}, {Wallerstein} \&
  {Mottini}}{{McWilliam} et~al.}{2013}]{McWilliam13}
{McWilliam} A.,  {Wallerstein} G.,    {Mottini} M.,  2013, \apj, 778, 149

\bibitem[\protect\citeauthoryear{{Monaco}, {Murante}, {Borgani} \&
  {Fontanot}}{{Monaco} et~al.}{2006}]{Monaco06}
{Monaco} P.,  {Murante} G.,  {Borgani} S.,    {Fontanot} F.,  2006, \apjl, 652,
  L89

\bibitem[\protect\citeauthoryear{{Muratov}, {Kere{\v s}},
  {Faucher-Gigu{\`e}re}, {Hopkins}, {Quataert} \& {Murray}}{{Muratov}
  et~al.}{2015}]{Muratov15}
{Muratov} A.~L.,  {Kere{\v s}} D.,  {Faucher-Gigu{\`e}re} C.-A.,  {Hopkins}
  P.~F.,  {Quataert} E.,    {Murray} N.,  2015, \mnras, 454, 2691

\bibitem[\protect\citeauthoryear{{Nagashima}, {Lacey}, {Okamoto}, {Baugh},
  {Frenk} \& {Cole}}{{Nagashima} et~al.}{2005}]{Nagashima05}
{Nagashima} M.,  {Lacey} C.~G.,  {Okamoto} T.,  {Baugh} C.~M.,  {Frenk} C.~S.,
    {Cole} S.,  2005, \mnras, 363, L31

\bibitem[\protect\citeauthoryear{{Narayanan} \& {Dav{\'e}}}{{Narayanan} \&
  {Dav{\'e}}}{2013}]{NarayananDave13}
{Narayanan} D.,  {Dav{\'e}} R.,  2013, \mnras, 436, 2892

\bibitem[\protect\citeauthoryear{{Padovani} \& {Matteucci}}{{Padovani} \&
  {Matteucci}}{1993}]{PadovaniMatteucci93}
{Padovani} P.,  {Matteucci} F.,  1993, \apj, 416, 26

\bibitem[\protect\citeauthoryear{{Papadopoulos}}{{Papadopoulos}}{2010}]{Papado%
poulos10}
{Papadopoulos} P.~P.,  2010, \apj, 720, 226

\bibitem[\protect\citeauthoryear{{Papadopoulos}, {Thi}, {Miniati} \&
  {Viti}}{{Papadopoulos} et~al.}{2011}]{Papadopoulos11}
{Papadopoulos} P.~P.,  {Thi} W.-F.,  {Miniati} F.,    {Viti} S.,  2011, \mnras,
  414, 1705

\bibitem[\protect\citeauthoryear{{Peeples}, {Werk}, {Tumlinson}, {Oppenheimer},
  {Prochaska}, {Katz} \& {Weinberg}}{{Peeples} et~al.}{2014}]{Peeples14}
{Peeples} M.~S.,  {Werk} J.~K.,  {Tumlinson} J.,  {Oppenheimer} B.~D.,
  {Prochaska} J.~X.,  {Katz} N.,    {Weinberg} D.~H.,  2014, \apj, 786, 54

\bibitem[\protect\citeauthoryear{{Pflamm-Altenburg}, {Weidner} \&
  {Kroupa}}{{Pflamm-Altenburg} et~al.}{2007}]{PflammAltenburg07}
{Pflamm-Altenburg} J.,  {Weidner} C.,    {Kroupa} P.,  2007, \apj, 671, 1550

\bibitem[\protect\citeauthoryear{{Pipino}, {Devriendt}, {Thomas}, {Silk} \&
  {Kaviraj}}{{Pipino} et~al.}{2009}]{Pipino09}
{Pipino} A.,  {Devriendt} J.~E.~G.,  {Thomas} D.,  {Silk} J.,    {Kaviraj} S.,
  2009, \aap, 505, 1075

\bibitem[\protect\citeauthoryear{{Pipino} \& {Matteucci}}{{Pipino} \&
  {Matteucci}}{2004}]{PipinoMatteucci04}
{Pipino} A.,  {Matteucci} F.,  2004, \mnras, 347, 968

\bibitem[\protect\citeauthoryear{{Planck Collaboration XVI}}{{Planck
  Collaboration XVI}}{2014}]{Planck_cosmpar}
{Planck Collaboration XVI} 2014, \aap, 571, A16

\bibitem[\protect\citeauthoryear{{Popping}, {Caputi}, {Trager}, {Somerville},
  {Dekel}, {Kassin}, {Kocevski}, {Koekemoer}, {Faber}, {Ferguson}, {Galametz},
  {Grogin}, {Guo}, {Lu}, {Wel} \& {Weiner}}{{Popping} et~al.}{2015}]{Popping15}
{Popping} G.,  {Caputi} K.~I.,  {Trager} S.~C.,  {Somerville} R.~S.,  {Dekel}
  A.,  {Kassin} S.~A.,  {Kocevski} D.~D.,  {Koekemoer} A.~M.,  {Faber} S.~M.,
  {Ferguson} H.~C.,  {Galametz} A.,  {Grogin} N.~A.,  {Guo} Y.,  {Lu} Y.,
  {Wel} A.~v.~d.,    {Weiner} B.~J.,  2015, \mnras, 454, 2258

\bibitem[\protect\citeauthoryear{{Pozzetti}, {Cimatti}, {Zamorani}, {Daddi},
  {Menci}, {Fontana}, {Renzini}, {Mignoli}, {Poli}, {Saracco}, {Broadhurst},
  {Cristiani}, {D'Odorico}, {Giallongo} \& {Gilmozzi}}{{Pozzetti}
  et~al.}{2003}]{Pozzetti03}
{Pozzetti} L.,  {Cimatti} A.,  {Zamorani} G.,  {Daddi} E.,  {Menci} N.,
  {Fontana} A.,  {Renzini} A.,  {Mignoli} M.,  {Poli} F.,  {Saracco} P.,
  {Broadhurst} T.,  {Cristiani} S.,  {D'Odorico} S.,  {Giallongo} E.,
  {Gilmozzi} R.,  2003, \aap, 402, 837

\bibitem[\protect\citeauthoryear{{Recchi}, {Calura} \& {Kroupa}}{{Recchi}
  et~al.}{2009}]{Recchi09}
{Recchi} S.,  {Calura} F.,    {Kroupa} P.,  2009, \aap, 499, 711

\bibitem[\protect\citeauthoryear{{Salpeter}}{{Salpeter}}{1955}]{Salpeter55}
{Salpeter} E.~E.,  1955, \apj, 121, 161

\bibitem[\protect\citeauthoryear{{S{\'a}nchez}, {Kennicutt}, {Gil de Paz}, {van
  de Ven}, {V{\'{\i}}lchez}, {Wisotzki}, {Walcher}, {Mast} \& et
  al.}{{S{\'a}nchez} et~al.}{2012}]{Sanchez12}
{S{\'a}nchez} S.~F.,  {Kennicutt} R.~C.,  {Gil de Paz} A.,  {van de Ven} G.,
  {V{\'{\i}}lchez} J.~M.,  {Wisotzki} L.,  {Walcher} C.~J.,  {Mast} D.,    et
  al. 2012, \aap, 538, A8

\bibitem[\protect\citeauthoryear{{Saracco}, {Fiano}, {Chincarini}, {Vanzella},
  {Longhetti}, {Cristiani}, {Fontana}, {Giallongo} \& {Nonino}}{{Saracco}
  et~al.}{2006}]{Saracco06}
{Saracco} P.,  {Fiano} A.,  {Chincarini} G.,  {Vanzella} E.,  {Longhetti} M.,
  {Cristiani} S.,  {Fontana} A.,  {Giallongo} E.,    {Nonino} M.,  2006,
  \mnras, 367, 349

\bibitem[\protect\citeauthoryear{{Segers}, {Schaye}, {Bower}, {Crain},
  {Schaller} \& {Theuns}}{{Segers} et~al.}{2016}]{Segers16}
{Segers} M.~C.,  {Schaye} J.,  {Bower} R.~G.,  {Crain} R.~A.,  {Schaller} M.,
   {Theuns} T.,  2016, \mnras, 461, L102

\bibitem[\protect\citeauthoryear{{Smith}}{{Smith}}{2014}]{Smith14}
{Smith} R.~J.,  2014, \mnras, 443, L69

\bibitem[\protect\citeauthoryear{{Smith}, {Lucey} \& {Conroy}}{{Smith}
  et~al.}{2015}]{Smith15}
{Smith} R.~J.,  {Lucey} J.~R.,    {Conroy} C.,  2015, \mnras, 449, 3441

\bibitem[\protect\citeauthoryear{{Spolaor}, {Kobayashi}, {Forbes}, {Couch} \&
  {Hau}}{{Spolaor} et~al.}{2010}]{Spolaor10}
{Spolaor} M.,  {Kobayashi} C.,  {Forbes} D.~A.,  {Couch} W.~J.,    {Hau}
  G.~K.~T.,  2010, \mnras, 408, 272

\bibitem[\protect\citeauthoryear{{Springel}, {White}, {Jenkins}, {Frenk},
  {Yoshida}, {Gao}, {Navarro}, {Thacker}, {Croton}, {Helly}, {Peacock}, {Cole},
  {Thomas}, {Couchman}, {Evrard}, {Colberg} \& {Pearce}}{{Springel}
  et~al.}{2005}]{Springel05}
{Springel} V.,  {White} S.~D.~M.,  {Jenkins} A.,  {Frenk} C.~S.,  {Yoshida} N.,
   {Gao} L.,  {Navarro} J.,  {Thacker} R.,  {Croton} D.,  {Helly} J.,
  {Peacock} J.~A.,  {Cole} S.,  {Thomas} P.,  {Couchman} H.,  {Evrard} A.,
  {Colberg} J.,    {Pearce} F.,  2005, \nat, 435, 629

\bibitem[\protect\citeauthoryear{{Tacconi}, {Genzel}, {Neri}, {Cox}, {Cooper},
  {Shapiro}, {Bolatto}, {Bouch{\'e}} \& et al.}{{Tacconi}
  et~al.}{2010}]{Tacconi10}
{Tacconi} L.~J.,  {Genzel} R.,  {Neri} R.,  {Cox} P.,  {Cooper} M.~C.,
  {Shapiro} K.,  {Bolatto} A.,  {Bouch{\'e}} N.,    et al. 2010, \nat, 463, 781

\bibitem[\protect\citeauthoryear{{Tacconi}, {Neri}, {Genzel}, {Combes},
  {Bolatto}, {Cooper}, {Wuyts} \& {Bournaud}}{{Tacconi}
  et~al.}{2013}]{Tacconi13}
{Tacconi} L.~J.,  {Neri} R.,  {Genzel} R.,  {Combes} F.,  {Bolatto} A.,
  {Cooper} M.~C.,  {Wuyts} S.,    {Bournaud} F. e.~a.,  2013, \apj, 768, 74

\bibitem[\protect\citeauthoryear{{Thielemann}, {Argast}, {Brachwitz}, {Hix},
  {H{\"o}flich}, {Liebend{\"o}rfer}, {Martinez-Pinedo}, {Mezzacappa}, {Nomoto}
  \& {Panov}}{{Thielemann} et~al.}{2003}]{Thielemann03}
{Thielemann} F.-K.,  {Argast} D.,  {Brachwitz} F.,  {Hix} W.~R.,  {H{\"o}flich}
  P.,  {Liebend{\"o}rfer} M.,  {Martinez-Pinedo} G.,  {Mezzacappa} A.,
  {Nomoto} K.,    {Panov} I.,  2003, in Hillebrandt W., Leibundgut B., eds,
  From Twilight to Highlight: The Physics of Supernovae Supernova
  Nucleosynthesis and Galactic Evolution. Springer-Verlag, Berlin, p.~311

\bibitem[\protect\citeauthoryear{{Thomas}, {Maraston}, {Bender} \& {Mendes de
  Oliveira}}{{Thomas} et~al.}{2005}]{Thomas05}
{Thomas} D.,  {Maraston} C.,  {Bender} R.,    {Mendes de Oliveira} C.,  2005,
  \apj, 621, 673

\bibitem[\protect\citeauthoryear{{Thomas}, {Maraston} \& {Johansson}}{{Thomas}
  et~al.}{2011}]{Thomas11a}
{Thomas} D.,  {Maraston} C.,    {Johansson} J.,  2011, \mnras, 412, 2183

\bibitem[\protect\citeauthoryear{{Thomas}, {Maraston}, {Schawinski}, {Sarzi} \&
  {Silk}}{{Thomas} et~al.}{2010}]{Thomas10}
{Thomas} D.,  {Maraston} C.,  {Schawinski} K.,  {Sarzi} M.,    {Silk} J.,
  2010, \mnras, 404, 1775

\bibitem[\protect\citeauthoryear{{Trager}, {Faber} \& {Dressler}}{{Trager}
  et~al.}{2008}]{Trager08}
{Trager} S.~C.,  {Faber} S.~M.,    {Dressler} A.,  2008, \mnras, 386, 715

\bibitem[\protect\citeauthoryear{{Trager}, {Faber}, {Worthey} \&
  {Gonz{\'a}lez}}{{Trager} et~al.}{2000}]{Trager00}
{Trager} S.~C.,  {Faber} S.~M.,  {Worthey} G.,    {Gonz{\'a}lez} J.~J.,  2000,
  \aj, 119, 1645

\bibitem[\protect\citeauthoryear{{Tremonti}, {Heckman}, {Kauffmann},
  {Brinchmann}, {Charlot}, {White}, {Seibert}, {Peng}, {Schlegel}, {Uomoto},
  {Fukugita} \& {Brinkmann}}{{Tremonti} et~al.}{2004}]{Tremonti04}
{Tremonti} C.~A.,  {Heckman} T.~M.,  {Kauffmann} G.,  {Brinchmann} J.,
  {Charlot} S.,  {White} S.~D.~M.,  {Seibert} M.,  {Peng} E.~W.,  {Schlegel}
  D.~J.,  {Uomoto} A.,  {Fukugita} M.,    {Brinkmann} J.,  2004, \apj, 613, 898

\bibitem[\protect\citeauthoryear{{Walcher}, {Wisotzki}, {Bekerait{\'e}},
  {Husemann}, {Iglesias-P{\'a}ramo}, {Backsmann}, {Barrera Ballesteros},
  {Catal{\'a}n-Torrecilla} \& et al.}{{Walcher} et~al.}{2014}]{Walcher14}
{Walcher} C.~J.,  {Wisotzki} L.,  {Bekerait{\'e}} S.,  {Husemann} B.,
  {Iglesias-P{\'a}ramo} J.,  {Backsmann} N.,  {Barrera Ballesteros} J.,
  {Catal{\'a}n-Torrecilla} C.,    et al. 2014, \aap, 569, A1

\bibitem[\protect\citeauthoryear{{Wang}, {De Lucia}, {Kitzbichler} \&
  {White}}{{Wang} et~al.}{2008}]{Wang08}
{Wang} J.,  {De Lucia} G.,  {Kitzbichler} M.~G.,    {White} S.~D.~M.,  2008,
  \mnras, 384, 1301

\bibitem[\protect\citeauthoryear{{Weidner} \& {Kroupa}}{{Weidner} \&
  {Kroupa}}{2005}]{WeidnerKroupa05}
{Weidner} C.,  {Kroupa} P.,  2005, \apj, 625, 754

\bibitem[\protect\citeauthoryear{{Weidner}, {Kroupa} \& {Larsen}}{{Weidner}
  et~al.}{2004}]{Weidner04}
{Weidner} C.,  {Kroupa} P.,    {Larsen} S.~S.,  2004, \mnras, 350, 1503

\bibitem[\protect\citeauthoryear{{Weidner}, {Kroupa} \&
  {Pflamm-Altenburg}}{{Weidner} et~al.}{2011}]{Weidner11}
{Weidner} C.,  {Kroupa} P.,    {Pflamm-Altenburg} J.,  2011, \mnras, 412, 979

\bibitem[\protect\citeauthoryear{{Weidner}, {Kroupa}, {Pflamm-Altenburg} \&
  {Vazdekis}}{{Weidner} et~al.}{2013}]{Weidner13}
{Weidner} C.,  {Kroupa} P.,  {Pflamm-Altenburg} J.,    {Vazdekis} A.,  2013,
  \mnras, 436, 3309

\bibitem[\protect\citeauthoryear{{Yates}, {Henriques}, {Thomas}, {Kauffmann},
  {Johansson} \& {White}}{{Yates} et~al.}{2013}]{Yates13}
{Yates} R.~M.,  {Henriques} B.,  {Thomas} P.~A.,  {Kauffmann} G.,  {Johansson}
  J.,    {White} S.~D.~M.,  2013, \mnras, 435, 3500

\bibitem[\protect\citeauthoryear{{Zibetti}, {Charlot} \& {Rix}}{{Zibetti}
  et~al.}{2009}]{Zibetti09}
{Zibetti} S.,  {Charlot} S.,    {Rix} H.,  2009, \mnras, 400, 1181

\end{thebibliography}
